\documentclass[11pt,twoside]{article}
\usepackage{epsf}
\input epsf.tex

\usepackage{amsmath}
\usepackage{amsfonts}
\usepackage{epsfig}
\usepackage{atmp04,psfrag}

\def \IR{\mathbb{R}}
\def \IC{\mathbb{C}}
\def \IZ{\mathbb{Z}}
\def \IP{\mathbb{P}}
\def \IB{\mathbb{B}}
\def \IF{\mathbb{F}}

\def \tx{{\tilde x}}
\def \ty{{\tilde y}}
\def \tu{{\tilde u}}
\def \tv{{\tilde v}}

\begin{document}
\setcounter{page}{459}

\newcommand{\Eqn}[1]{\hspace{-.5em}& #1 & \hspace{-.5em}}

\newcommand{\onefigure}[2]{\begin{figure}[h]
         \caption{\small #2\label{#1}(#1)}
         \end{figure}}
\newcommand{\onefigurenocap}[1]{\begin{figure}[h]
         \begin{center}\leavevmode\epsfbox{#1.eps}\end{center}
         \end{figure}}
\renewcommand{\onefigure}[2]{\begin{figure}[h]
         \begin{center}\leavevmode\epsfbox{#1.eps}\end{center}
         \caption{\small #2\label{#1}}
         \end{figure}}
\newcommand{\comment}[1]{}
\newcommand{\myref}[1]{(\ref{#1})}
\newcommand{\secref}[1]{sec.~\protect\ref{#1}}
\newcommand{\figref}[1]{Fig.~\protect\ref{#1}}
\newcommand{\mathbold}[1]{\mbox{\boldmath $\bf#1$}}
\newcommand{\mJ}{\mathbold{J}}
\newcommand{\momega}{\mathbold{\omega}}
\newcommand{\bz}{{\bf z}}
\def\bbbz{{\sf Z\!\!\!Z}}
\newcommand{\PP}{\mbox{I}\!\mbox{P}}
\newcommand{\ff}{\mbox{I}\!\mbox{F}}
\newcommand{\bbbc}{\mbox{C}\!\!\!\mbox{I}}
\def\sl2z{SL(2,\bbbz)}
\newcommand{\bbbq}{I\!\!Q}
\newcommand{\be}{\begin{equation}}
\newcommand{\ee}{\end{equation}}
\newcommand{\bea}{\begin{eqnarray}}
\newcommand{\eea}{\end{eqnarray}}
\newcommand{\nn}{\nonumber}
\newcommand{\unit}{1\!\!1}
\newcommand{\half}{\frac{1}{2}}
\newcommand{\shalf}{\mbox{$\half$}}
\newcommand{\transform}[1]{
   \stackrel{#1}{-\hspace{-1.2ex}-\hspace{-1.2ex}\longrightarrow}}
\newcommand{\inter}[2]{\null^{\#}(#1\cdot#2)}
\newcommand{\lprod}[2]{\vec{#1}\cdot\vec{#2}}
\newcommand{\mult}[1]{{\cal N}(#1)}
\newcommand{\Bn}{{\cal B}_N}
\newcommand{\B}{{\cal B}}
\newcommand{\Beight}{{\cal B}_8}
\newcommand{\Bnine}{{\cal B}_9}
\newcommand{\Eman}{\widehat{\cal E}_N}
\newcommand{\C}{{\cal C}}
\newcommand{\Q}{Q\!\!\!Q}
\newcommand{\comp}{C\!\!\!C}
\newdimen\tableauside\tableauside=1.0ex
\newdimen\tableaurule\tableaurule=0.4pt
\newdimen\tableaustep
\def\phantomhrule#1{\hbox{\vbox to0pt{\hrule height\tableaurule width#1\vss}}}
\def\phantomvrule#1{\vbox{\hbox to0pt{\vrule width\tableaurule height#1\hss}}}
\def\sqr{\vbox{%
  \phantomhrule\tableaustep
  \hbox{\phantomvrule\tableaustep\kern\tableaustep\phantomvrule\tableaustep}%
  \hbox{\vbox{\phantomhrule\tableauside}\kern-\tableaurule}}}
\def\squares#1{\hbox{\count0=#1\noindent\loop\sqr
  \advance\count0 by-1 \ifnum\count0>0\repeat}}
\def\tableau#1{\vcenter{\offinterlineskip
  \tableaustep=\tableauside\advance\tableaustep by-\tableaurule
  \kern\normallineskip\hbox
    {\kern\normallineskip\vbox
      {\gettableau#1 0 }%
     \kern\normallineskip\kern\tableaurule}%
  \kern\normallineskip\kern\tableaurule}}
\def\gettableau#1 {\ifnum#1=0\let\next=\null\else
  \squares{#1}\let\next=\gettableau\fi\next}

\tableauside=1.0ex
\tableaurule=0.4pt

\def\IE{\relax{\rm I\kern-.18em E}}
\def\IP{\relax{\rm I\kern-.18em P}}

\noindent

\copyrightnotice{2003}{7}{459}{499}

\title{Instanton Counting and Chern-Simons Theory}
\author{Amer Iqbal$^{1}$, Amir-Kian Kashani-Poor$^{2}$}
\address{$^{1}$ Jefferson Laboratory,
Harvard University,\\
Cambridge, MA  02138, U.S.A.}
\address{$^{2}$ Department of Physics and SLAC,
Stanford University,\\
Stanford, CA 94305/94309, U.S.A.}
\arxurl{hep-th/0212279}
\barefootnote{HUTP-02/A064;SU-ITP 02/48;SLAC-PUB-9623}
\pagestyle{myheadings}
\markboth{\it Instanton Counting and Chern-Simons Theory}{\it A. Iqbal
  and A. Kashani-Poor}
\begin{abstract}
The instanton partition function of ${\cal N}=2$, 
$D=4$ $SU(2)$ gauge theory is obtained by taking 
the field theory limit of the topological open 
string partition function, given by a Chern-Simons 
theory, of a CY3-fold. The CY3-fold on the open 
string side is obtained by geometric transition from 
local $\IP^{1}\times \IP^{1}$ which is used in 
the geometric engineering of the $SU(2)$ theory. The 
partition function obtained from the Chern-Simons 
theory agrees with the closed topological string 
partition function of local $\IP^{1}\times \IP^{1}$ 
proposed recently by Nekrasov. We also obtain the 
partition functions for local $\IF_{1}$ and $\IF_2$ CY3-folds and 
show that the topological string amplitudes of all three local Hirzebruch surfaces give rise to the same field theory limit. It is 
shown that a generalization of the topological closed 
string partition function whose field theory limit is 
the generalization of the instanton partition function, 
proposed by Nekrasov, can be determined easily from the 
Chern-Simons theory.
\end{abstract}


\section{Introduction}
Large N dualities in the context of closed and open 
topological strings on different CY3-fold backgrounds 
have been the source of much excitement recently \cite{DV1,DV2}. 
These dualities have interesting consequences for 
both ${\cal N}=2$ and ${\cal N}=1$ D=4 field theories 
which can be geometrically engineered using the 
type II strings and D-branes on the CY3-folds \cite{KLMVW,KKV,CV1,CV2}. One 
example of large N duality, which will be relevant 
for our purpose, is the calculation of the partition function
of A-model topological closed strings propagating on a 
CY3-fold from the partition function of topological 
open strings on a different CY3-fold \cite{GV1,AMV,DFG}. The CY3-fold on 
which the open strings propagate is obtained from the 
CY3-fold which is the background of the closed topological 
strings by multiple conifold-like transitions on the 
exceptional curves \cite{GV1}. The open string theory on the dual 
CY3-fold reduces to a Chern-Simons theory on each of 
the $S^{3}$'s \cite{WCS}, obtained by the transition 
from exceptional curves, plus corrections coming from 
holomorphic curves with boundaries on the 3-cycles 
\cite{AMV,DFG}. 

The A-model topological string amplitude (the $\log$ of the 
topological closed string partition function) is the generating function of 
Gromow-Witten invariants of all genera and therefore 
is the answer to an enumerative problem \cite{BCOV}. It also has 
a physically interesting interpretation in the 
${\cal N}=2$ D=4 theory obtained by compactifying 
type IIA strings on a CY3-fold: the topological string 
amplitude gives certain holomorphic corrections to 
the effective action of the four dimensional theory 
\cite{BCOV}. In the context of geometric engineering
of gauge theories the genus zero amplitude, in a 
certain limit, computes both the perturbative 
and instanton corrections to the prepotential of the
${\cal N}=2$ gauge theory \cite{KLMVW,KKV}.

In this paper we show that it is possible to obtain 
the exact instanton partition function \cite{hollowood,Nekrasov, algorithm,BFMT} 
\footnote{By instanton partition function, ${\cal Z}(\hbar)$, we mean the field theory
limit (see section \ref{geo}) of the topological string partition function $\sum g_s^{2g-2} F_g$ such that the prepotential of the field theory
is given by $\lim_{\hbar\rightarrow 0}\,\hbar^{2}\log{\cal Z}(\hbar)$. An intrinsic field theory
definition of this is given in terms of the topological twisted four dimensional theory \cite{Nekrasov}.} of the 
${\cal N}=2$ $SU(2)$ SYM \cite{SW1,SW2} by taking the field theory 
limit of an open string partition function. The 
topological open strings propagate on a CY3-fold which
is obtained by multiple geometric transitions from the 
local $\IF_{m}$ CY3-folds used in the geometric 
engineering of the ${\cal N}=2$ SU(2) SYM theory 
\cite{KKV,KLMVW}. The instanton partition function
obtained in this way agrees exactly with the partition 
function proposed recently by Nekrasov \cite{Nekrasov} and calculated in \cite{BFMT}.
Moreover, the complete open string partition function 
agrees with the A-model partition function of local 
$\IF_{0}$ obtained by Nekrasov from an index 
calculation \cite{Nekrasov}. A more general 
partition function can be obtained by taking the
Chern-Simons coupling constant to be different for 
different 3-cycles as opposed to the usual 
identification of the Chern-Simons coupling constants
with the string coupling constant, 
$\frac{2\pi}{k_{i}+N_{i}}=g_{s}$. The field theory
limit of this partition function agrees with the
generalized instanton partition functions proposed by 
Nekrasov \cite{Nekrasov, algorithm}.

The paper is organized as follows. In section two we 
briefly review the geometric engineering of 
${\cal N}=2$ SU(2) SYM theory from local 
$\IF_{m}$ CY3-folds. In section three, we review the geometric transitions at the heart of the open-closed large N-duality \cite{AMV}. We consider the case of local $\IP^2$ in detail, and use these results to motivate the expected transitions for local $\IF_{m}$ 3-folds. In section four, we evaluate the Chern-Simons partition functions for the local Hirzebruch surfaces. We obtain the partition function in a form that is well-suited for taking the field theory limit. In section 
five, we show that the field theory limit of the 
partition function gives an exact expression for 
the instanton partition function which agrees with 
the expression given by Nekrasov and is the same 
for all local $\IF_{m}$. We also show that the full 
partition function, after some rearrangement of the 
factors, is exactly equal to the A-model expression 
given by Nekrasov \cite{Nekrasov}. In the appendix, we fill in some details on the geometric transition in the case of local $\IP^2$ and give some curve counting functions which can be 
used to determine integer invariants of three
and four instanton contribution to higher genus 
corrections.

\begin{figure}[h]
\psfrag{closed}{Topological closed strings}
\psfrag{closedtwo}{(A-model) on local $\IF_m$}
\psfrag{open}{Topological open strings}
\psfrag{opentwo}{on deformation of local $\IF_m$}
\psfrag{sutwo}{${\cal N}=2$ $SU(2)$ SYM}
\psfrag{cs}{Chern-Simons theories}
\psfrag{cstwo}{with Wilson line insertions}
\psfrag{large}{large N-duality}
\psfrag{field}{$\alpha' \rightarrow 0$}
\psfrag{sectiontwo}{section 2}
\psfrag{sectionthree}{section 3}
\psfrag{sectionfour}{section 4}
\psfrag{sectionfive}{section 5}
\psfrag{world}{{\small world-sheet instantons}}
\psfrag{space}{{\small space-time instantons}}
\psfrag{worldwith}{{\small world-sheet instantons}}
\psfrag{worldwithtwo}{{\small with boundary}}
\begin{flushleft}\epsfxsize=.85\textwidth\leavevmode\epsfbox{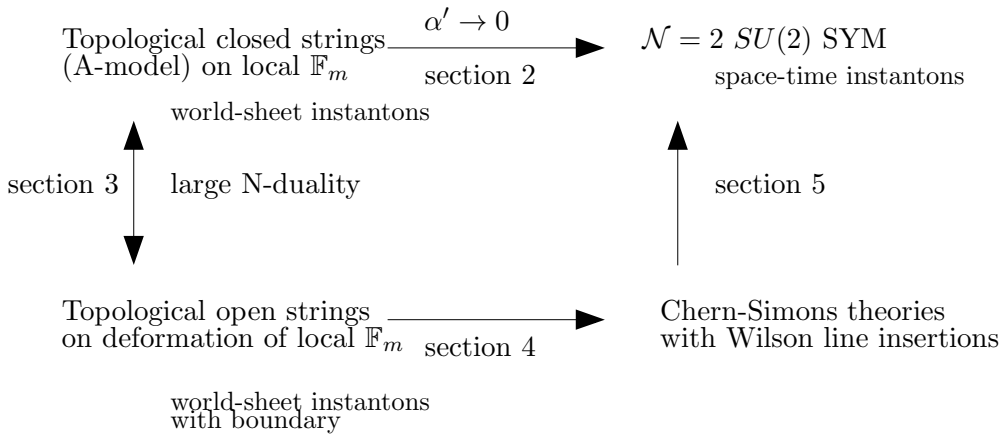}\end{flushleft} 
\caption{\small Outline of paper.}
\end{figure}

\section{Geometrical engineering of pure ${\cal N}=2$ 
$SU(2)$ theory} \label{geo}

The Calabi-Yau threefold compactification of type IIA 
strings provides a very powerful way of studying 
${\cal N}=2$ D=4 quantum field theories \cite{KLMVW,KKV}. 
The prepotential of the D=4 theory is given by 
the genus zero topological string amplitude of the 
corresponding Calabi-Yau threefold. The gauge symmetry 
in the field theory arises from D2-branes wrapping
collapsing curves in the CY3-fold. Thus to get a 
particular gauge symmetry one has to study a 
CY3-fold with the appropriate singularity.
We will restrict ourselves to the case of $SU(2)$ 
gauge symmetry and only discuss the relevant CY3-folds for this case.

Geometrical engineering of pure ${\cal N}=2$ 
theories with $SU(2)$ gauge symmetry was studied
in \cite{KLMVW,KKV}. The relevant singularity is of $A_1$ type, i.e. of the form $\IC^2/\IZ_2$. The local CY2-fold $T^* \IP^1$ develops this type of singularity as we take the area of the base to zero. To obtain an effective 4 dimensional theory, we need to fiber this space over an additional $\IP^1$. We hence choose a Hirzebruch surface $\IF_{m}$ as the compact base of the local CY3-fold. The line bundle over this base which leads to a total space of vanishing first Chern class is the canonical line bundle. For $m >2$, the total space of the canonical line bundle contains additional compact 4-cycles, aside from $\IF_m$. 
Although for this reason we will restrict attention to $\IF_{0}$, $\IF_{1}$ and  $\IF_{2}$, our results hold for general $m$ as well. For $m>2$, the partition function derived in the next section allows us to obtain the invariants of those curves lying
in $\IF_{m}$.

Let us now briefly review the field theory limit. To obtain this limit, we push the string scale to infinity. By asymptotic freedom of the 4 dimensional gauge theory, the gauge coupling hence goes to zero. Since the 6 dimensional and the 4 dimensional gauge coupling are related via the area of the base $\IP^1$ of the Hirzebruch surface, the field theory limit requires large base size. At the same time, to keep the mass of the W-bosons, given by the area of the fiber (remember that the gauge symmetry enhancement occurs when this area shrinks to zero), finite in the limit in which the string scale is taken to infinity, we must consider the small area limit of the fiber. Since the running of the gauge coupling is dominated at weak coupling by the logarithm of the W-boson mass, these two limits are related as $T_B \sim - \log T_F$, where $T_B$ and $T_F$ denote the K\"ahler parameters corresponding to the base and the fiber of the Hirzebruch surface.\footnote{Here $T_B,T_F$ 
are the quantum corrected K\"ahler parameters i.e., 
they are solutions of the relevant Picard-Fuchs 
equations that go like $T_{b,f}=t_{b,f}+
\sum_{n,m}c_{n,m}e^{-nt_{b}-mt_{f}}$, where 
$t_{b,f}$ are the K\"ahler parameters occurring in the 
linear sigma model description of local $\IF_{m}$.}
In fact, by invoking the discrete symmetry of the gauge theory (left over from the anomalous $U(1)$ R-charge), we know that the $n$-instanton contribution $\sim e^{-\frac{n}{g^2}}$ is accompanied by a factor $(\frac{1}{a})^{4n}$ \cite{Seiberg}, where $a$ parametrizes the VEV of the scalar field breaking the $SU(2)$ gauge symmetry, $a \sim T_F$. Retaining all instanton contributions in the field theory limit hence requires scaling the K\"ahler parameters as
\begin{eqnarray} \label{fieldtheorylimit}
Q_{B}:=e^{-T_B}=(\frac{\beta \Lambda}{2})^{4}\,,\,\,\,Q_{F}:=e^{-T_F}=e^{-2\beta a}\,. 
\end{eqnarray}
Here, $\Lambda$ is the quantum scale in four 
dimensions, and the parameter $\beta$ is introduced such that the field theory limit corresponds to $\beta\rightarrow 0$. 
In section \ref{curves}, we will be taking the field theory limit of the topological partition function $\sum g_s^{2g-2} F_g$. It turns out that we obtain finite contributions from all genera if we scale the string coupling such that $q:=e^{i g_s}=e^{\beta \hbar}$. $\hbar$ will serve to distinguish between the contributions at different $g_s$ (the notation is chosen in accordance with \cite{Nekrasov}).

The prepotential of the theory
has both 1-loop perturbative and non-perturbative (instanton)
contributions,
\bea
{\cal F}={\cal F}_{classical}+{\cal F}_{1-loop}+a^{2}\sum_{k=1}^{\infty}
c_{k}(\frac{\Lambda}{a})^{4k}\,.
\eea
The $k$-instanton contribution to the prepotential, $c_{k}(\frac{\Lambda}{a})^{4k}$, comes from world-sheet instantons 
wrapping the curves $\{kB+mF~|~m=0,1,2,\cdots\}$ and 
is therefore captured by the field theory limit 
of the genus zero topological string amplitude \cite{KKV}.
From the expansion of the genus zero topological 
string amplitude
\bea
F_{0}(T_B,T_F)&=&P_{3}(T_B,T_F)+\sum_{(k,m)\neq (0,0)}
\sum_{n=1}^{\infty}\frac{N^{0}_{(k,m)}}{n^{3}}
e^{-nkT_B-nmT_F}\,,
\eea
(here $P_{3}(T_B,T_F)$ is a cubic polynomial from which one gets the classical
contribution to the prepotential) 
it is clear that the $k$-instanton contribution is 
proportional to the regularized sum 
$\sum_{m}N^{0}_{(k,m)}$, where $N^{0}_{(k,m)}$ is 
(up to a sign) the Euler characteristic of the 
moduli space of the D-brane wrapped on the curve 
$kB+mF$. In section \ref{curves}, we will see that these 
sums can be easily obtained from the Chern-Simons theory 
arising in the open string geometry dual to the 
closed string geometry of the CY3-fold used in the 
geometric engineering of the $SU(2)$ theory. 

There are various ways of obtaining the closed 
topological string amplitudes: localization, 
B-model calculations, or large N duality with topological open 
strings.  Direct localization calculations are 
difficult since we want to sum up the contribution 
of all curves $kB+mF$ for a fixed $k$, B-model 
calculations can sum up the contribution of 
all curves to the k-th instanton sector, as was 
discussed in detail in \cite{KMT}, but become more 
difficult for large $k$ and for higher genus. 
We will therefore use the large N duality with topological open 
strings on a deformed CY3-fold background to 
determine the exact instanton partition function. This method yields all higher genus contributions to the closed string partition function simultaneously. More precisely, the all genus closed string free energy was shown in \cite{GV} to have the following integrality structure
\begin{eqnarray} \label{gvpfsin}
F_{closed}(\omega) :=\sum_{g=0}^{\infty}g_{s}^{2g-2}F_{g}(\omega)=\sum_{\Sigma\in H_{2}(X)}\sum_{g=0}^{\infty}\sum_{n=1}^{\infty}
\frac{{N}^{g}_{\Sigma}}{n} (2\sin(n\frac{ g_s}{2}))^{2g-2}\,e^{-n\Sigma\cdot \omega}\,.
\end{eqnarray}
The Chern-Simons calculation yields all Gopakumar-Vafa invariants ${N}^{g}_{\Sigma}$ up to a given degree in $\Sigma$. 

\section{Closed to open transition}
Open topological string theory on a local CY $X$ is related to CS-theory in the following way \cite{WCS}: a CS theory lives on every Lagrangian submanifold of $X$ on which open strings can end. In addition, contributions from strings wrapping compact holomorphic curves in $X$ and ending on these submanifolds are captured in the CS theory by insertions of 
Wilson lines: these compute the holonomy of the CS-connection around the boundaries of the holomorphic curves. Since the $U(N)$ gauge bundles over the Lagrangian submanifolds are required to be flat, these Wilson lines calculate invariants of the homotopy class of the curves.

Enumerating the compact holomorphic curves of a complex manifold is usually a very difficult problem. The crucial ingredient in calculating closed world-sheet instantons using Chern-Simons theory following the methods of \cite{AMV} consists in deforming the local Calabi-Yau which is the target space of the closed topological string to obtain a geometry in which the compact holomorphic curves are under strict control: they are isolated cylinders and their multicovers, stretching between certain of the Lagrangian submanifolds. The basic local model for this deformation is the conifold transition, which we briefly review.

On the closed string side, one considers the bundle ${\cal O}(-1) \oplus {\cal O}(-1) \rightarrow \IP^1$. Taking the volume of the $\IP^1$ to $0$ yields the singular geometry of the conifold. This space is described by the equation $xy=uv$ in $\IC^4$. String theory on this space is not singular if we turn on the NS-NS 2-form. This setup is described by a purely imaginary complexified K\"ahler parameter $t$ of the $\IP^1$, $t=\frac{2\pi iN}{k+N}$ (this choice will not be a limitation on what we can compute on the closed string side, as the partition functions are holomorphic in $t$, i.e. can be obtained for arbitrary value of $t$ by analytic continuation). The singular geometry allows a deformation, described by $xy=uv+\mu$, $\mu \in \IC$, which replaces the singular locus of the conifold by an $S^3$ with volume $\mu$. Since $\mu$ is a complex structure moduli, the $A$-model amplitudes we are considering do not depend on it. 

\begin{figure}[h]
\begin{center}\epsfxsize=.8\textwidth\leavevmode\epsfbox{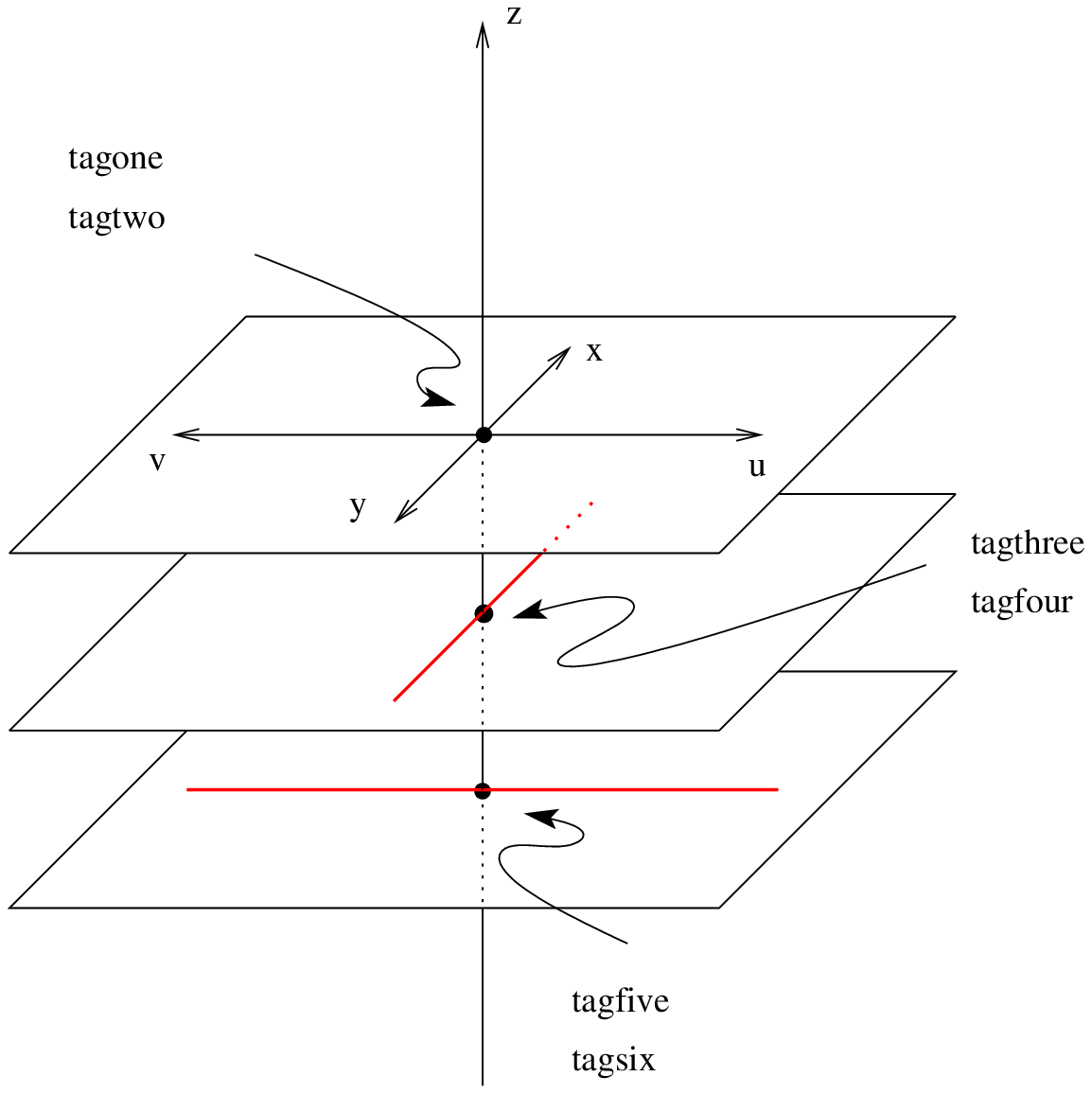}\end{center} 
\caption{\small Deformed conifold; a cycle of the $T^2$ fiber degenerates along each of the red lines (the horizontal, vertical line respectively in the two lower planes). \label{liftinglines}}
\end{figure}

If we introduce an additional $\IC$ valued variable $z$, s.t. $z=xy$ and $z=uv +\mu$, we can visualize the deformed geometry as a $\IR^2 \times T^2$ fibration over $\IC$ as follows. At each value of $z$, we have a real plane. One real axis is parametrized by gluing the two half-lines $|x| \in [\sqrt{|z|},\infty)$ and  $|y| \in [\sqrt{|z|},\infty)$ at $|x|=|y|=\sqrt{|z|}$, the other analogously for $u$ and $v$. As far as these real planes are concerned, nothing special happens at $z=0$ and $z=\mu$. This is not true for the $T^2$ factor of the fiber. The compact $T^2$ is coordinatized by the phases of $x$, $y$ and $u$, $v$, with the transition functions $\phi_x=-\phi_y$, $\phi_u=-\phi_v$ on the overlaps at $|x|=|y|$, $|u|=|v|$ respectively. We see that a cycle degenerates along a line in the real plane at the values $z=0$ and $z=\mu$.
This geometry is encoded in \figref{liftinglines}.
\psfrag{tagone}{$(|x|=|y|=\sqrt{|z|},$}
\psfrag{tagtwo}{$|u|=|v|=\sqrt{|z-\mu|},z)$}
\psfrag{tagthree}{$(|x|=|y|=\sqrt{|\mu|},$}
\psfrag{tagfour}{$|u|=|v|=0,z=\mu)$}
\psfrag{tagfive}{$(|x|=|y|=0,$}
\psfrag{tagsix}{$|u|=|v|=\sqrt{|\mu|},z=0)$}
\psfrag{x}{$|x|$}
\psfrag{y}{$|y|$}
\psfrag{u}{$|u|$}
\psfrag{v}{$|v|$}
\psfrag{z}{Re($z$)}

The $S^3$ of the deformed conifold is given by the $T^2$ fibration over the real line $z=t, x=y= \sqrt{t}, u=v=\sqrt{t-\mu}$ for $t\in [0,\mu]$ which connects the two degeneration loci. Note that the only holomorphic curves in this geometry ending on $S^3$, i.e. intersecting the $S^3$ along a circle, have constant $z$ value. This will be an important constraint in finding the compact holomorphic curves in related geometries with several Lagrangian 3-manifolds, to which we now turn.

\subsection{Local $\IP^2$}

\begin{figure}[h]
\begin{center}\epsfxsize=\textwidth\leavevmode\epsfbox{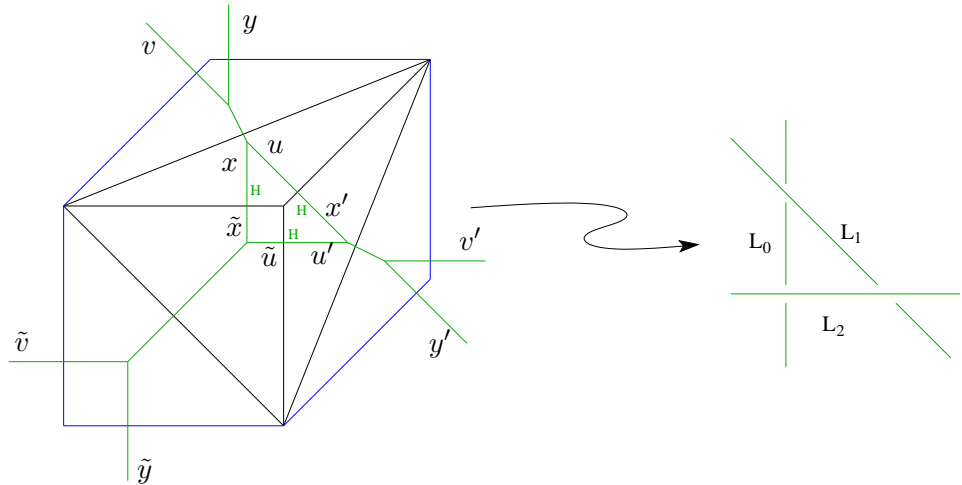}\end{center} 
\caption{\small Deforming flopped local $\IB_3$. The lines $L_0$, $L_1$, $L_2$ encode the degeneration locus of the $T^2$ fiber in the real plane at $z=0,\mu_1,\mu_2$ respectively. \label{deformingb3}}
\end{figure}

As a concrete example, consider ${\cal O}(-3) \rightarrow \IP^2$. Though the base contains three $\IP^1$s invariant under the torus action (these are given by $(a:b:0), (a:0:b), (0:a:b)$), they are in the same Weyl class as the hypersurface divisor, i.e. are not exceptional and hence cannot undergo a conifold transition. To obtain a manageable geometry for the CS theory, i.e. a geometry of the type referred to above, we blow up the three toric fixed points of the base, obtaining a local del Pezzo $\IB_3$.
Using standard methods, detailed in the appendix, we find three patches with which we can cover the singular limit of this geometry in which the size of the three exceptional curves is taken to 0. Each patch is described by 4 variables, $x,y,u,v$ in the first patch, the corresponding variables tilded, primed in the second and third patch, satisfying one constraint equation, $xy = uv$ in the first patch, the tilded, primed version of this equation in the second, third patch. The transition functions between these patches are given in the appendix. The three exceptional divisors we have obtained in this way can undergo a conifold transition. We perform the following deformations
\begin{eqnarray}
xy &=& uv + \mu_1 \,,\\
\tx \ty  &=& \tu \tv + \mu_2 \,,\\
x' y' + \mu_1 &=& u' v' + \mu_2 \,. 
\end{eqnarray}
The resulting geometry is depicted in \figref{deformingb3}).

\psfrag{px}{$x$}
\psfrag{py}{$y$}
\psfrag{pu}{$u$}
\psfrag{pv}{$v$}
\psfrag{tx}{$\tx$}
\psfrag{ty}{$\ty$}
\psfrag{tu}{$\tu$}
\psfrag{tv}{$\tv$}
\psfrag{xp}{$x'$}
\psfrag{yp}{$y'$}
\psfrag{up}{$u'$}
\psfrag{vp}{$v'$}

As in the conifold case, we now introduce a new variable $z$ such that $z = xy =\tx \ty = x' y' + \mu_1$. For this system of equations to be consistent, we must also deform the transition functions between the patches. This is done in the appendix. An important feature of this construction is that the relation between the phases of the complex coordinates in the overlap of the different patches, as given by the transition functions, is encoded in the relative slopes of the lines $L_0,L_1,L_2$ in \figref{deformingb3}. As the two cycles of the $T^2$ fiber are given by the phases of these coordinates, the slopes of these lines allow us to read off which cycle of the $T^2$ is degenerating in the real planes at $z=0$, $z=\mu_1$, and $z=\mu_2$.

Lines in the complex $z$-plane connecting the points $z=0$, $z=\mu_1$, $z=\mu_2$ have a $T^2$ fibered over them with a cycle degenerating at either end. Each of these $T^2$ fibrations over an interval form a closed 3-manifold, which has a convenient description, with regard to the CS evaluation \cite{Witten:1988hf,AMV}, in terms of a Heegaard splitting \cite{dubrovin}. This splitting allows us to describe the 3-manifolds as obtained by gluing two solid tori together along their surface using an $SL(2,\IZ)$ map ($SL(2,\IZ)$ encompasses all self-diffeomorphisms of a torus up to homotopy). To this end, we divide the interval $I$ connecting the two points in the $z$-plane into two intervals $I_1$, $I_2$. $I_1$ and $I_2$ coordinatize the depth direction of the two solid tori. The torus worth of points at each depth $i \in I_1$ or $i \in I_2$ is identified with the fiber over the corresponding point of the interval $I$. The two endpoints of $I$, over which the $T^2$ fiber degenerates, correspond to the depth 0 points $\partial I_1 \cap \partial I$ or $\partial I_2 \cap \partial I$ of the solid tori. To obtain a correct realization of the $T^2$ fibration over $I$, we must in the final step glue the two solid tori together such that the appropriate cycles are identified. Let us introduce a basis for the second homology of the three $T^2$s in the game, $A$ and $B$ for the $T^2$ fiber over $I$, such that the cycles degenerating at the endpoints of $I$ are $A$ and $aA+bB$, $A_i$ and $B_i$, $i=1,2$ for the tori comprising the surfaces of the two solid tori, such that $A_1$ and $A_2$ are the cycles which become trivial when we fill in the tori. Hence, we need to identify $A$ with $A_1$ and $aA+bB$ with $A_2$. This fixes two of the entries of the $SL(2,\IZ)$ gluing diffeomorphism between the surfaces of the two solid tori. Next, we would like to fix the $B_i$ cycles to be the $S^1$s along which the two compact holomorphic annuli intersect the 3-manifold. Note that since the $A_i$ cycles of the fiber are degenerate along the holomorphic curves, multiples of these can be added to the $B_i$ at will. Hence, the above identifications fix the $SL(2,\IZ)$ gluing diffeomorphism only up to this ambiguity,
\begin{eqnarray} \label{gluingdif}
\left(\begin{array}{cc}
     1 & 0 \\
     1 & 1 
\end{array} \right)^{n_2}
\left(\begin{array}{c}
      A_2 \\ B_2
\end{array} \right)
=\left(\begin{array}{cc}
     a & b \\
     c & d 
\end{array} \right)
\left(\begin{array}{cc}
     1 & 0 \\
     1 & 1 
\end{array} \right)^{n_1}
\left(\begin{array}{c}
    A_1 \\ B_1
\end{array} \right) \,,
\end{eqnarray}
where $n_1$ and $n_2$ are arbitrary integers. In the CS framework, the cycles $B_i+n_i A_i$ are wrapped by Wilson loops. The integer ambiguity $(n_1,n_2)$ in the choice of these matrices corresponds to a framing ambiguity in the CS picture which we will fix by hand.

It is not hard to show that the three 3-manifolds arising in our geometry are all $S^3$s. \footnote{This is done by an analysis of the first fundamental group of the 3-manifold obtained by gluing the two solid tori \cite{dubrovin}. Let us briefly sketch how information on the 3-manifold $M$ can be obtained from considering $\pi_1(M)$. By the Seifert-Van Kampen theorem, this group is generated by the disjoint union of the generators of the two tori modulo the relations imposed by the gluing diffeomorphism. In the notation above, these two generators are $B_1$ and $B_2$ (or more precisely the images of these cycles under the embedding of the surface tori into the solid tori). By equation (\ref{gluingdif}) and the triviality of the embedded cycles $A_1$ and $A_2$, $B_1^b=1$. Hence, if $|b| \neq 0,1$, $\pi_1(M)$ contains a torsion element. In particular, $M$ cannot be $S^3$. A more careful analysis \cite{dubrovin} shows that the only 3-manifolds one can obtain via the construction described above are $S^3$, $S^2 \times S^1$, and Lens spaces $L_b$.} After the deformation, we hence arrive at an $S^3$ situated at each vertex of the original web diagram. This geometry contains compact holomorphic curves. These curves end on the $S^3$s. The same considerations as in the conifold case show that any such curve must therefore have constant $z$ coordinate. By choosing the complex deformation parameters appropriately, we can ensure that the $S^3$s pairwise intersect only in one point in the $z$-plane (recall that $z$ is a complex coordinate), and that these intersection points all coincide with values of $z$ at which some cycle of the $T^2$ fibration degenerates (in other words, the finite intervals which represent the $S^3$s in the $z$-plane only touch at their endpoints). Arguing patchwise, we can easily see that the only compact holomorphic curves in the geometry thus obtained (annuli and their multicovers) have axes along the line in the $\IR^2$ plane along which a cycle on the $T^2$ degenerates: in the first patch, say, at $z=0$, we must satisfy $xy=0, uv=-\mu_1$ and furthermore, we want to intersect the $S^3$ at $x=\bar{y}$, $u=-\bar{v}$ in a circle. It is not hard to show that this is only possible for $x=y=0, u=s^m$ for $s \in \IC$, $|s| \ge |\mu_1|^{\frac{1}{2m}}$ or $|s|\le |\mu_1|^{\frac{1}{2m}}$.

We must establish the relation between the open and closed geometry parameters: the complex K\"ahler parameters $t_i$ of the blown down exceptional $\IP^1$s of the closed string geometry are the 't Hooft couplings on the open string side, $t_i= \frac{2\pi i N_i}{k+N_i}$. In addition, we have the K\"ahler parameters $r'_i$ of the curves in the base that do not partake in the conifold transition. On the open string, the corresponding parameters classically should be the areas of the world-sheet instanton annuli $r_i$ stretched between the $S^3$s (the geometric data on the open string side also includes the volumes of the $S^3$s, which are however complex structure moduli and therefore not relevant for the A-model amplitudes). To relate the CS-partition function we obtain on the geometry of \figref{deformingb3} to the closed string partition function of local $\IP^2$, we must consider the limit $t_i \rightarrow \infty$ (again, we are deforming K\"ahler parameters; the reason we can do this without impunity is that the closed string partition functions are holomorphic in these parameters).
For this limit to exist, we will see that the K\"ahler parameters $r'_i$ of the non-exceptional curves of the closed string geometry must receive contributions, in the mapping from open to closed string parameters, from the 't Hooft couplings in addition to the expected contribution from the area $r_i$ of the annuli. It would be interesting to see this more directly, e.g., as suggested in \cite{AMV}, by utilizing the GL$\sigma$M approach employed in \cite{OV} to prove the large N-duality in the conifold case.

From this example, the path we would like to follow for a local CY on any toric base is clear: we would like to blow up the vertices of the toric fan (these correspond to fixed points of the torus action), then perform a conifold transition on each of the exceptional curves so as to obtain a 3-manifold at each such vertex. \footnote{That this deformation is always possible is suggested by analogues of \figref{deformingb3}. However, since we argue patchwise, care needs to be taken that the deformed patches can consistently be glued together. This will require deforming the transition functions in addition to the constraint equations in each patch.} This gives rise to Feynman-like rules for computing the closed string partition function, as pointed out in \cite{Amer}.

\subsection{Hirzebruch surfaces}

\begin{figure}[h]
\begin{center}\epsfxsize=\textwidth\leavevmode\epsfbox{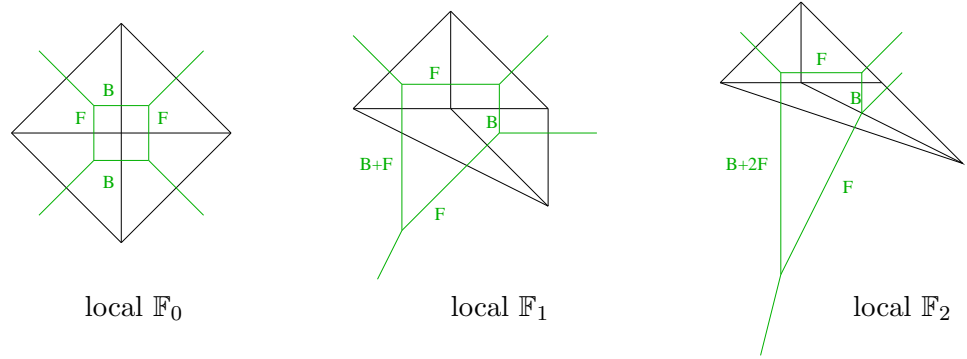}\end{center} 
\caption{\small The fans and web diagrams for the canonical line bundle over the first three Hirzebruch surfaces. \label{fansandwebs}}
\end{figure}

The geometries that will be relevant for the field theory application in this paper are the canonical line bundles over the first three Hirzebruch surfaces $\IF_m$ \footnote{for $m>2$ the surface $\IF_{m}$ is accompanied by other 4-cycles in the non-compact CY3-fold. For these cases, the partition functions that we will write down give only the contribution of curves in $\IF_{m}$ to 
the total partition function.} . These surfaces are $\IP^1$ bundles over a $\IP^1$ base. $\IF_0$ is the trivial bundle $\IP^1 \times \IP^1$. The toric fans and web diagrams for these local CYs are depicted in \figref{fansandwebs}.

\psfrag{tagfzero}{local $\IF_0$}
\psfrag{tagfone}{local $\IF_1$}
\psfrag{tagftwo}{local $\IF_2$}

The second homology $H_{2}(\IF_{m},\IZ)$ of the Hirzebruch surfaces is spanned by the cycles $B$ and $F$, represented by the base and fiber. The intersection numbers of these cycles are
\bea
B^{2}=-m\,,\,F^{2}=0\,,\,\,B\cdot F=1\,.
\eea

\begin{figure}[h]
\begin{flushleft}\epsfxsize=\textwidth\leavevmode\epsfbox{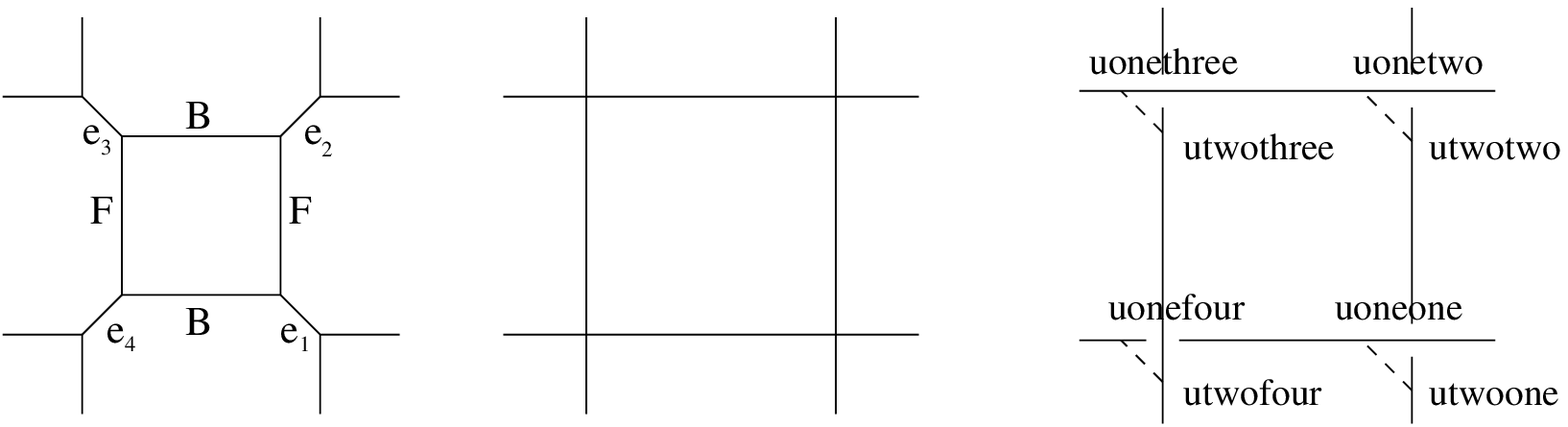}\end{flushleft} 
\caption{\small The large N transition in the case of $\IF_{0}$. The $U^i_j$ indicate Wilson loop insertions in the CS theories as explained in section \ref{cspf}.}\label{f0}
\end{figure}

To obtain an open string geometry, we blow up the vertices of these diagrams and perform the conifold transition on the exceptional divisors thus obtained.\footnote{Note that the base of $\IF_1$ is an exceptional curve. In fact, $\IF_1$ and $\IB_1$, i.e. $\IP^{2}$ blown up at one point, are isomorphic. Hence, to calculate the partition function of $\IF_1$, we could perform the conifold transition on this exceptional curve and only two additional ones obtained by blowing up the two vertices at which $F$ and $B+F$ intersect in \figref{fansandwebs}. This yields exactly the geometry of local $\IB_3$ that we considered in the case of local $\IP^2$. To regain $\IF_1$, we now would send two of the three K\"ahler paramters $t_i$ of the exceptional curves to infinity \cite{AMV}. We can even obtain the partition function for $\IF_2$ from this geometry, by sending the appropriate combination of K\"ahler parameters to infinity.} This is illustrated diagrammatically for the case of $\IF_{0}$ in \figref{f0}.
 
\psfrag{uoneone}{$U^1_1$}
\psfrag{utwoone}{$U^2_1$}
\psfrag{uonetwo}{$U^1_2$}
\psfrag{utwotwo}{$U^2_2$}
\psfrag{uonethree}{$U^1_3$}
\psfrag{utwothree}{$U^2_3$}
\psfrag{uonefour}{$U^1_4$}
\psfrag{utwofour}{$U^2_4$}

Note that in blowing up local $\IF_2$, the fan of the geometry is no longer convex. In the web diagram, this manifests itself in terms of crossing lines. The local CY hence contains additional 4-cycles. It would be interesting to check explicitly whether this affects the deformation argument. The fact that we obtain the correct invariants using the open geometry we naively obtain from such a deformation suggests that this is not so.

\section{Chern-Simons partition function for local Hirzebruch surfaces} \label{cspf}

In this section, we compute the CS partition functions $Z_{CS}$, with the appropriate Wilson loop insertions, based on the geometries obtained from deforming the local Hirzebruch surfaces $\IF_m$, $m=0,1,2$ as outlined in the previous section. We will follow closely the discussion of \cite{AMV} where the case of local $\IF_{0}$ was discussed in detail. According to the large N duality conjecture, these computations should reproduce the closed topological string partition functions (\ref{gvpfsin}) for the respective local CY3-folds. We will use this equality in the next section to determine the Gopakumar-Vafa invariants of the closed geometries as well as to study the field theory limit of the compactification of type IIA on them.

We will start by discussing the case $\IF_0$ which was studied in detail in \cite{AMV}. The open string geometry is depicted in \figref{f0}.
The Chern-Simons partition function for the full geometry is given by \cite{AMV}
\bea \label{CSpartitionfunction}
Z_{CS}(r_{i},N_{i};q)&=&\int \prod_{i=1}^{4}({\cal
  D}A_{i}e^{S_{CS}(A_{i})}){\cal O}(U^2_{1},U^1_{2};r_{B}){\cal
  O}(U^2_{2},U^1_{3};r_{F})\nonumber\\ & &\hspace{1em}{\cal O}(U^2_{3},U^1_{4};r_{B}){\cal O}(U^2_{4},U^1_{1};r_{F}).
\eea
Here, $\log {\cal O}(U,V;r)$ is the correction to the Chern-Simons 
action coming from annuli of length $r$ with boundary on two $S^{3}$'s \cite{AMV}(note that we have taken $r_{1}=r_{3}=r_B$ and $r_{2}=r_{4}=r_F$ in equation (\ref{CSpartitionfunction}) in accordance with the $\IF_{0}$ geometry),
\bea \label{instcorr}
{\cal O}(U ,V;r)=\mbox{exp}\{\sum_{n=1}^{\infty}\frac{e^{-nr}}{n}\mbox{Tr}U^{n}\mbox{Tr}V^{-n}\}=\sum_{R}e^{-l_{R}r}\mbox{Tr}_{R}U\,\mbox{Tr}_{R}V^{-1}\,.
\eea
$U^i_j$ computes the holonomy of the CS-connection $A_j$ along the boundary $\gamma_i$ of a compact holomorphic annulus ending on the $j$-th $S^3$ (in the cases we consider, there will be two curves $\gamma_1$ and $\gamma_2$ per $S^3$),
\bea
U^i_j = \mbox{Pexp}\oint_{\gamma_i} A_j \,.
\eea
The last equality in equation (\ref{instcorr}) follows from an application of the Frobenius formula. The sum is over all representations of the special unitary group, $l_R$ counts the number of boxes in the Young tableaux of the representation $R$.
This identity allows us to write the partition function (\ref{CSpartitionfunction}) as a simple sum of products of partition functions of the individual CS-theories,
\bea
Z_{CS}(r_{i},N_{i};q)&=&\langle{\cal O}(U^2_{1},U^1_{2};r_{B}){\cal
  O}(U^2_{2},U^1_{2};r_{F}){\cal O}(U^2_{3},U^1_{4};r_{B}){\cal
  O}(U^2_{4},U^1_{1};r_{F}) \rangle\,,\nn \\ 
&=&\sum_{R_{1,2,3,4}}e^{-r_{B}(l_{1}+l_{3})-r_{F}(l_{2}+l_{4})}W_{R_{1}R_{4}}(\lambda_{4},q)W_{R_{4}R_{3}}(\lambda_{3},q)\nonumber\\
& &\hspace{2em} W_{R_{3}R_{2}}(\lambda_{2},q)W_{R_{2}R_{1}}(\lambda_{1},q)\,.
\eea
$W_{R_{i}R_{j}}$ are expectation values in the individual CS theories with 't Hooft coupling $t_i = \frac{2 \pi i N_i}{k_i + N_i} = \log \lambda_i$ (recall that $q=e^{i g_s}$, hence $\lambda_i = q^{N_i}$), given by
\bea
W_{R_{i}R_{j}}(\lambda,q)=\langle \mbox{Tr}_{R_{i}}(U^1)\mbox{Tr}_{R_{j}}(U^2)\rangle \;.
\eea
In the following, we will reserve the notation $W_{R_{i}R_{j}}$ for the special constellation of curves $\gamma_i$ that occurs in the case of $\IF_0$ in each of the four $S^3$s, at zero framing (i.e. $n_1=n_2=0$, in the notation of equation (\ref{gluingdif})): the $\gamma_i$ wrap orthogonal cycles, hence form a Hopf link. The Hopf link invariants $W_{R_{i}R_{j}}$ can be easily calculated using the results of \cite{ML,lukac}.

As explained in the previous section, we need to consider the limit $\lambda_i \rightarrow \infty$ in order to recover the partition function for $\IF_0$, and later for $\IF_1$ and $\IF_2$. The leading power of $\lambda$ in $W_{R_{1}R_{2}}(\lambda,q)$ is $\lambda^{(l_{R_{1}}+l_{R_{2}})/2}$ \cite{AMV}. Hence, $Z_{CS}$ naively diverges in this limit. We remedy this by scaling $r_i$ together with $\lambda_i$, such that the appropriate linear combination of these parameters is finite in the $\lambda_i \rightarrow \infty$ limit. We interpret these linear combinations as the renormalized K\"ahler parameters of the closed string geometry. They are given by
\bea
T_B = r_B + \frac{t_2 + t_3}{2} =  r_B + \frac{t_1 + t_4}{2}  \,, \\
T_F = r_F + \frac{t_1 + t_2}{2} =  r_F + \frac{t_3 + t_4}{2} \,.
\eea
The CS partition function for local $\IF_{0}$ now becomes 
\bea\nn 
Z_{CS}(Q_B,Q_F;q)&=&\sum_{R_{1,2,3,4}}Q_B^{-(l_{1}+l_{3})}
Q_F^{-(l_{2}+l_{4})} {\cal W}_{R_{1}R_{4}}(q)\,{\cal
  W}_{R_{4}R_{3}}(q)\nonumber\\ & &\hspace{2em}{\cal W}_{R_{3}R_{2}}(q){\cal W}_{R_{2}R_{1}}(q)\,, \label{ZCS}
\eea
where, as before, $Q_{B}=e^{-T_B},Q_{F}=e^{-T_F}$, and we have defined 
\bea
{\cal W}_{R_{i}R_{j}}(q)=\lim_{\lambda\rightarrow \infty}
\lambda^{-(l_{i}+l_{j})/2}W_{R_{i}R_{j}}(\lambda,q)\,.
\eea
We can simplify this expression and perform two of the sums over representations explicitly. To this end, we introduce the quantity
\bea \label{defK}
K_{R_{1}R_{2}}(Q)=\sum_{R}Q^{l_{R}}{\cal W}_{R_{1}R}(q)\,{\cal W}_{RR_{2}}(q)\,.
\eea
In terms of $K_{R_{1}R_{2}}(Q)$, the CS partition function in equation (\ref{ZCS}) becomes
\bea 
\label{Zm}
Z_{CS}(Q_{B},Q_{F};q)=
\sum_{R_{1},R_{2}}Q_{B}^{l_{1}+l_{2}}K_{R_{1}R_{2}}(Q_{F})^{2}\,.
\eea
We will denote the trivial representation by a point. From our discussion in the previous section, we see that the function $K_{\cdot \cdot}$ yields the partition function of the closed string on the local CY $T^*(\IP^1)\times \IC$. Below, based on the explicit evaluation of this case in \cite{AMV}, we will make an ansatz for the form of $K_{R_{1}R_{2}}(Q)$ in the case of arbitrary $R_1$ and $R_2$ which will drastically simplify the computation of this expression.

\begin{figure}[h]
\begin{center}\epsfxsize=.3\textwidth\leavevmode\epsfbox{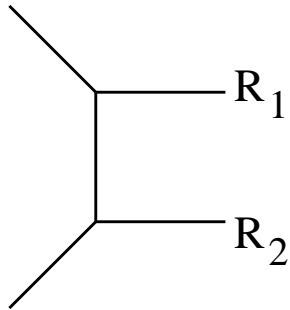}\end{center}
\caption{\small $K_{R_{1}R_{2}}(Q)$}\label{K}
\end{figure}

Diagrammatically, we can depict $K_{R_{1}R_{2}}(Q)$ as in \figref{K}.

\begin{figure}[h]
         \begin{center}\epsfxsize=\textwidth\leavevmode\epsfbox{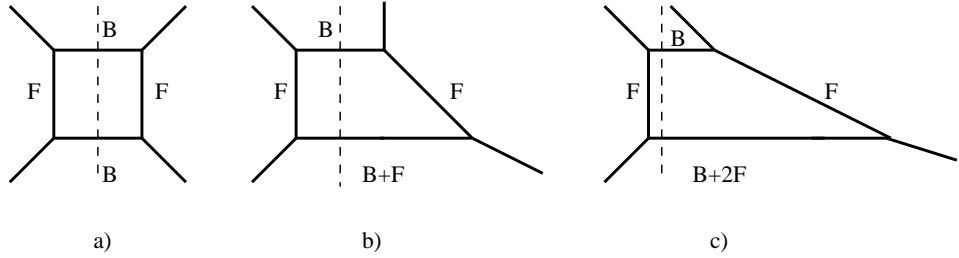}\end{center}
         \caption{\small Splitting local $\IF_m$ into $K^{(m)}_{R_i R_j}$ contributions.\label{F012}}
         \end{figure}

\begin{figure}[h]
         \begin{center}\epsfxsize=.4\textwidth\leavevmode\epsfbox{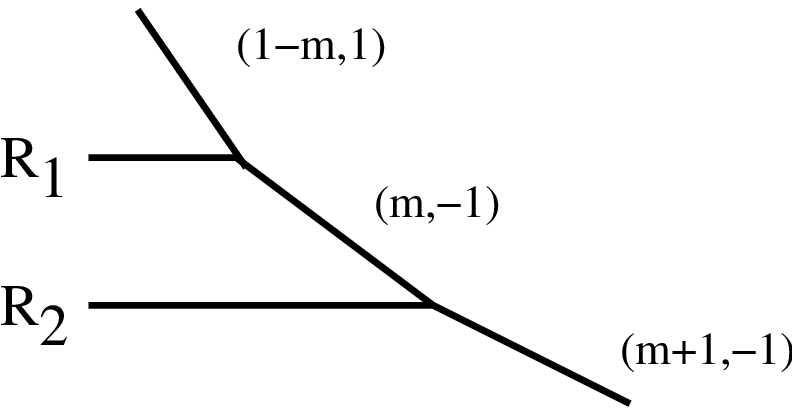}\end{center}
         \caption{\small $K^{(m)}_{R_1 R_2}$\label{slice}}
         \end{figure}

To evaluate the CS partition function for local $\IF_{1}$ and local $\IF_{2}$, we need similar expressions for the diagrams to the right of the dashed lines in \figref{F012}, as depicted in \figref{slice}. If we denote these contributions by $K^{(m)}_{R_{1}R_{2}}(Q)$, we can express the partition function for local $\IF_{(m)}$ as
\bea
Z^{(m)}_{CS}(Q_{B},Q_{F};q)=\sum_{R_{1},R_{2}}Q_{B}^{l_{R_{1}}+l_{R_{2}}}Q_{F}^{ml_{R_{2}}}K_{R_{1}R_{2}}(Q_{F})K^{(m)}_{R_{1}R_{2}}(Q_{F})\,,
\eea
where the factor of $Q_{B}^{ml_{R_{2}}}$ appears since the rational curve associated with the two 
parallel internal lines of the web diagram are $B$ and $B+mF$ for local $\IF_{m}$ (see \figref{fansandwebs}).

\begin{figure}[h]
         \begin{center}\epsfxsize=.4\textwidth\leavevmode\epsfbox{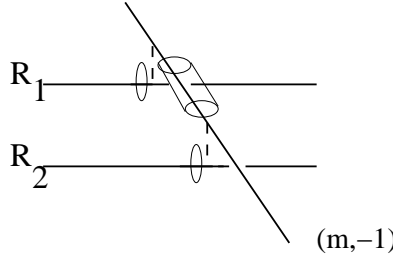}\end{center}
         \caption{\small The open string geometry which determined
  $K^{(m)}_{R_{1}R_{2}}(Q)$.\label{Km}}
         \end{figure}

The open string geometry which can be used to determine $K^{(m)}_{R_{1}R_{2}}(Q)$ is shown in \figref{Km} and is 
given by
\bea
K^{(m)}_{R_{1}R_{2}}(Q)&=& \lim_{\lambda\rightarrow \infty}\lambda^{l_R+\frac{l_{R_1}+l_{R_2}}{2}} \langle \mbox{Tr}_{R_{1}}U_{1}\,{\cal O}(U_{3},U_{4};r)\,
\mbox{Tr}_{R_{2}}U_{2}^{-1}\rangle\,,\\ \nn
&=& \lim_{\lambda\rightarrow \infty} \sum_{R} Q^{l_{R}}
\lambda^{\frac{l_{R_1}+l_{R}}{2}} W_{R_{1}R}(\lambda,q)
\lambda^{\frac{l_{R}+l_{R_2}}{2}} W_{RR_{2}}(\lambda,q)\,\nn\\ && \hspace{2em}
(-1)^{m(l_{R_{1}}+l_{R_{2}})}q^{\frac{m}{2}(\kappa_{R_{2}}-\kappa_{R_{1}})}\,,\nn
\eea
where $Q=e^{-r}$. The $m$ dependent factors stem from a choice of framing $n_1=-n_2=m$, $n_1=-n_2=-m$ at the two vertices respectively. \footnote{In the notation of \cite{AMV}, this corresponds to the choice of gluing matrices $T^{-m}S^{-1}T^{-m}$ and $T^m S^{-1} T^m$.} This choice was made by hand by matching the lowest Gopakumar-Vafa invariants we obtain with those calculated in the literature using localization methods. It would clearly be desirable to justify this choice intrinsically.

Comparing to equation (\ref{defK}), we obtain
\bea
K^{(m)}_{R_{1}R_{2}}(Q)=(-1)^{m(l_{R_{1}}+l_{R_{2}})}q^{\frac{m}{2}(\kappa_{R_{2}}-\kappa_{R_{1}})}K_{R_{1}R_{2}}(Q)\,.
\eea
The CS partition function in equation (\ref{Zm}) now becomes
\bea
Z^{(m)}_{CS}(Q_{B},Q_{F};q)&=&\sum_{R_{1},R_{2}}Q_{B}^{l_{R_{1}}+l_{R_{2}}}Q_{F}^{ml_{R_{2}}}(-1)^{m(l_{R_{1}}+l_{R_{2}})}\nn\\
& &\hspace{2em}q^{\frac{m}{2}(\kappa_{R_{2}}-\kappa_{R_{1}})}K_{R_{1}R_{2}}(Q_{F})^{2}\,.
\eea
In order to evaluate $K_{R_{1}R_{2}}(Q)$, we need to calculate the
leading order contribution in $\lambda$ to $W_{R_{1}R_{2}}(\lambda,q)$. We include the necessary formulae for completeness. For a derivation of the following relations, we refer the reader to the cited references and the recent review article \cite{marino}.

The leading order contribution to $W_{R_{1}R_{2}}(\lambda,q)$ is given by \cite{ML,lukac,AMV}
\bea
{\cal W}_{R_{1}R_{2}}(q)={\cal W}_{R_{1}}(q)\,q^{l_{R_{2}}/2}\,S_{\mu^{2}}(E_{\mu^{1}}(t))\,.
\eea
Here, $\mu^1$ and $\mu^2$ are the Young tableaux associated to the representations $R_1$ and $R_2$. $S_\mu$ is the polynomial $S_\mu = \det M_\mu$, where the $r \times r$ matrix $M_\mu$, $r$ being the number of columns in $\mu$, is given by $M_\mu^{(ij)}=(a_{\mu^{\vee}_{i}+j-i})$. $\mu^{\vee}$ denotes the transposed Young tableaux to $\mu$, i.e. with columns and rows interchanged. The $a_i$ are the coefficients of $t^i$ in the expansion of $E_{\mu}$, given by
\bea
E_{\mu}(t)=(1+\sum_{n=1}^{\infty}(\prod_{i=1}^{n}\frac{1}{q^{i}-1})t^{n})\,(\prod_{j=1}^{d}\frac{1+q^{\mu_{j}-j}t}{1+q^{-j}t})\,,
\eea
\bea
{\cal W}_{R}(q)&=&q^{\kappa_{R}/4}\prod_{1\leq i<j\leq d}\frac{[\mu_{i}-\mu_{j}+j-i]}{[j-i]}\,\prod_{i=1}^{d}\prod_{v=1}^{\mu_{i}}\frac{1}{[v-i+d]}\,,
\label{weq}
\eea
where $[x]=q^{x/2}-q^{-x/2}$, $d$ denotes the number of rows in the tableau $\mu$, $\mu_i$ denotes the number of boxes in the $i$-th row of $\mu$, and $\kappa_R$ is given by
\begin{equation}
\kappa_R=l_R + \sum_{i=1}^{d(\mu)}\mu_i (\mu_i - 2i) \,.
\end{equation}

In the next section, we will take the field theory limit to get the instanton partition function. For this reason, it is important to know how ${\cal W}_{R}(q)$ behaves in the limit $q\rightarrow 1$. Since $[x] \approx (q-1)x$ in this limit, the first product on the RHS of equation (\ref{weq}) is finite in this limit, but the 
second diverges as $(q-1)^{-l_{R}}$. It follows that 
\bea \label{wl}
{\cal W}_{R}(q) = \frac{{\cal L}_{R}(q)}{(q-1)^{l_{R}}}\,,
\eea
where ${\cal L}_{R}(q)$ is finite at $q=1$ and given by
\bea
{\cal L}_{R}(1)=\prod_{1\leq i<j\leq d}\frac{\mu_{i}-\mu_{j}+j-i}{j-i}\,\prod_{i=1}^{d}\prod_{v=1}^{\mu_{i}}\frac{1}{v-i+d} \,.
\eea

If both $R_{1}$ and $R_{2}$ are the trivial representation, then, as pointed out above, $K_{R_{1} R_{2}}$ computes the closed string partition function for the CY3-fold $T^{*}(\IP^{1})\times \IC$. The partition function for this geometry was obtained in \cite{AMV} and has the form
\bea
K_{\cdot \,\cdot}(Q)=\mbox{Exp}\{\sum_{n=1}^{\infty}B_{n}(q)Q^{n}\}\,,\,\,B_{n}(q)=\frac{q^{n}}{n(q^{n}-1)^{2}}=\frac{B_{1}(q^{n})}{n}\,.
\eea
Notice that $K_{\cdot \cdot}(Q)$ diverges in the field theory limit. For non-trivial $R_{1},R_{2}$, we can parametrize $K_{R_{1} R_{2}}$ in the form 
\bea \label{krr}
K_{R_{1}R_{2}}(Q)=
K_{\cdot \cdot}(Q){\cal W}_{R_{1}}(q){\cal W}_{R_{2}}(q)\mbox{Exp}\{\sum_{n=1}^{\infty}f_{R_{1}R_{2}}^{n}(q)Q^{n}\}\,.
\eea
It is natural to expect that
\bea
f_{R_{1}R_{2}}^{n}(q)=\frac{f_{R_{1}R_{2}}(q^{n})}{n}\,,
\eea
even for non-trivial representations (the intuition behind this ansatz is that the $\frac{f(q^n)}{n}$ $n$-dependence stems from having contributions from a single isolated curve and its multicovers, which is a generic feature of the geometries to which the $K^{(m)}_{R_1 R_2}$ contribute). We have explicitly calculated the first few terms of $K_{R_{1}R_{2}}$ for some non-trivial 
representations and found the results to be in agreement with this ansatz. Using this ansatz, calculating the sum over all representations in $K_{R_{1}R_{2}}$ is reduced to determining a single term in the series, the coefficient of $Q$. This can easily be determined to be
\bea
f_{R_{1}R_{2}}(q)&=&\frac{{\cal W}_{R_{1},\tableau{1}}}{{\cal W}_{R_{1}}}
\frac{{\cal W}_{\tableau{1},R_{2}}}{{\cal W}_{R_{2}}}-{\cal W}_{\tableau{1}}^{2}\,\\
&=&
\frac{q}{(q-1)^{2}}\{1+(q-1)\sum_{j=1}^{d_{1}}(q^{\mu^1_{j}-j}-q^{-j})\}\nn\\
& &\hspace{2em}\{1+(q-1)\sum_{j=1}^{d_{2}}(q^{\mu^2_{j}-j}-q^{-j})\}-\frac{q}{(q-1)^{2}}\,\nn.
\eea
The above expression for $f_{R_{1}R_{2}}(q)$ can be simplified to the following form,
\bea \label{frr}
f_{R_{1}R_{2}}(q)&=&(q-2+q^{-1})f_{R_{1}}(q)f_{R_{2}}(q)+f_{R_{1}}(q)+f_{R_{2}}(q)\nn\\
&=:&
\sum_{k}C_{k}(R_{1},R_{2})q^{k}\,,
\eea
where
\bea \label{fr}
f_{R}(q):=f_{R,\,.}(q)&=&\sum_{j=1}^{d}q^{-(j-1)}(1+q+q^{2}+\cdots+q^{\mu_{j}-1})\,,\\ \nn
&=&\sum_{j=1}^{d}\sum_{v=1}^{\mu_{j}}q^{v-j}\,.
\eea
The following properties of the function $f_{R_{1}R_{2}}(q)$, which are easily read off from equations (\ref{frr}) and (\ref{fr}), will be useful later,
\bea
f_{R_{1}^{T}R_{2}^{T}}(q)&=&f_{R_{1}R_{2}}(q^{-1})\,,\\ \nn
f_{R_{1}R_{2}}(1)&=&\sum_{k}C_{k}(R_{1},R_{2})=l_{R_{1}}+l_{R_{2}}\,,\\ \nn
\left. \frac{df_{R_{1}R_{2}}(q)}{dq}\right\vert_{q=1}&=&\sum_{k}k\,C_{k}(R_{1},R_{2})=\frac{\kappa_{R_{1}}+\kappa_{R_{2}}}{2}\,.
\eea
Using the above form of $f_{R_{1}R_{2}}(q)$, we get
\bea
K_{R_{1}R_{2}}(Q)&=&K_{\cdot \cdot}(Q){\cal W}_{R_{1}}{\cal W}_{R_{2}}
\prod_{k}(1-q^{k}Q)^{-C_{k}(R_{1},R_{2})}\,,
\eea
and thus, the partition function is given by
\bea
Z^{(m)}_{CS}&=&K^{2}_{\cdot \cdot}(Q_{F})\sum_{R_{1},R_{2}}Q_{B}^{l_{R_{1}}+l_{R_{2}}}Q_{F}^{ml_{R_{2}}}
(-1)^{m(l_{R_{1}}+l_{R_{2}})}\nn\\ & &\hspace{2em}q^{\frac{m}{2}(\kappa_{R_{1}}-\kappa_{R_{2}})}
\frac{{\cal W}^{2}_{R_{1}}(q){\cal W}^{2}_{R_{2}}(q)}
{\prod_{k}(1-q^{k}Q_{F})^{2C_{k}(R_{1},R_{2})}}\,.
\label{pf}
\eea
Notice that, as a consequence of having performed two of the sums over representations in $Z^{(m)}_{CS}$ explicitly, the above expression allows us to compute, at every order in $Q_B$, the dependence on $Q_F$ to {\em all} orders. At given $n$, we thus obtain the Gopakumar-Vafa invariants $N^{g}_{(n,m)}$ for all $m$.

\section{Counting curves and instantons} \label{curves}

We now want to use the results obtained in the previous section to compute the Gopakumar-Vafa invariants of the local Hirzebruch surfaces and to study the field theory limit of type IIA string theory compactified on these spaces.

The closed string partition function for a CY 3-fold $X$, equation (\ref{gvpfsin}), can be put in the following form
\bea
F_{closed}(\omega)=\sum_{g=0}^{\infty}g_{s}^{2g-2}F_{g}(\omega)=\sum_{\Sigma\in H_{2}(X)}\sum_{g=0}^{\infty}\sum_{n=1}^{\infty}
\frac{\widehat{N}^{g}_{\Sigma}\,q^{n(1-g)}}{n(q^{n}-1)^{2-2g}}\,e^{-n\Sigma\cdot \omega}\,,
\eea
where $q=e^{ig_{s}}$ and $\omega$ is the quantum corrected K\"ahler form on $X$.  The integer invariants $\widehat{N}^{g}_{\Sigma}$ are essentially Gopakumar-Vafa invariants, related to the invariants $N^{g}_{\Sigma}$ introduced in \cite{GV} by $\widehat{N}^{g}_{\Sigma}=(-1)^{g-1}N^{g}_{\Sigma}$. Recall that $H_{2}(\IF_{m},\IZ)$ is spanned by the homology classes $B$ and $F$ of the base $\PP^{1}$ and the fiber $\PP^{1}$ respectively. The exponential in the partition function hence takes the form $e^{-n (k T_B + l T_F)}=Q_B^{n k} Q_F^{nl}$.       The generating functions of the topological string 
amplitudes defined above can then be written as
\bea
F_{closed}(T_B,T_F)&=&\sum_{n=1}^{\infty}\frac{\widehat{N}^{0}_{F}q^{n}}{n(q^{n}-1)^2}\,Q_{F}^{n}\nn\\
& &\mbox{}+\sum_{k=1}^{\infty}Q_{B}^{k}\sum_{g=0}^{\infty}\sum_{r|k}\frac{q^{r(1-g)}}{r(q^{r}-1)^{2-2g}}\,f^{(k/r)}_{g}(Q_{F}^{r})\,,
\label{eq1}
\eea
where 
\bea \label{gfgv}
f_{g}^{(n)}(x)=\sum_{m}\widehat{N}^{g}_{(n,m)}x^{m}\,.
\eea
$f^{(k)}_{g}(x)$ is the generating function for the genus $g$ invariants of curves $kB+*F$.
In the field theory limit, the first term in equation (\ref{eq1}), coming from multicovers of $F$,
gives the perturbative contribution to the prepotential as was shown in \cite{KKV}. We will
see that computations in Chern-Simons theory on the open string side determine the generating
functions of invariants $f^{(k)}_{g}(x)$ in a straightforward way. 

In subsection \ref{ftl}, we will take the field theory limit of the CS partition function directly, i.e. without the intermediate step of expressing it in terms of A-model quantities. We will see that, after a slight manipulation, it is the same as the expression given by Nekrasov \cite{Nekrasov}.

\subsection{Curves}
To extract the generating functions (\ref{gfgv}) for the Gopakumar-Vafa invariants from the CS partition functions (\ref{pf}), we need to equate the appropriate coefficients of $Q_B$ in $Z_{closed}=e^{F_{closed}}$ and $Z_{CS}$. To this end, we introduce the functions $G_k$ and $Z_k$, such that
\bea
Z_{closed}&=&\mbox{Exp}\{\sum_{n=1}^{\infty}\frac{G(q^{n},n\omega)}{n}\}\\ \nn
&=&K_{\cdot\,\cdot}(Q_{F})^{\widehat{N}^{0}_{(0,1)}}\,\mbox{Exp}\{\sum_{n=1}^{\infty}\frac{1}{n}\sum_{k=1}^{\infty}Q_{B}^{kn}G_{k}(q^{n},Q_{F}^{n})\} \,,
\eea
and
\bea
Z_{CS}(Q_{B},Q_{F};q)=K_{\cdot \,\cdot}^{2}(Q_{F})\,\sum_{k=0}^{\infty}Q_{B}^{k}Z_{k}(Q_{F},q)\,.
\eea
Then
\bea
G(q,\omega)&=&\sum_{\Sigma\in H_{2}(X,\bbbz)}\sum_{g=0}^{\infty}
\frac{\widehat{N}^{g}_{\Sigma}\,q^{1-g}}{(q-1)^{2-2g}}e^{-\Sigma\cdot \omega}\,\\\nn
&=&\frac{\widehat{N}^{0}_{(0,1)}\,q}{(q-1)^2}\,Q_{F}+
\sum_{k=1}^{\infty}Q_{B}^{k}G_{k}(q,Q_{F})\,,
\eea
where
\bea
G_{k}(q,Q_{F})=\sum_{m=0}^{\infty}\sum_{g=0}^{\infty}
\frac{\widehat{N}^{g}_{(k,m)}\,q^{1-g}}{(q-1)^{2-2g}}\,Q_{F}^{m}=\sum_{g=0}^{\infty}\frac{1}{(q^{1/2}-q^{-1/2})^{2-2g}}\,f_{g}^{(k)}(Q_{F})\,.
\eea
We have used the fact that $N^g_{(0,m)} \sim \delta_{g,0} \delta_{m,1}$, which was already pointed out in \cite{KKV}.\footnote{\cite{KKV} only considered the case $g=0$. The $\delta_{g,0}$ reflects the fact that a higher genus surface cannot be mapped holomorphically into a sphere.}
For $Z_k$, specializing to $\IF_0$ (the cases $\IF_1$ and  $\IF_2$ can be treated analogously; the results are 
quoted in the appendix), we obtain from equation (\ref{pf})
\bea
Z_{k}(q,Q_{F})=\sum_{\{R_1,R_2|l_{R_1}+l_{R_2}=k\}}\frac{{\cal W}_{R_{1}}^{2}\,
{\cal W}_{R_{2}}^{2}}{\prod_{m}(1-q^{m}Q_{F})^{2C_{m}(R_{1},R_{2})}}\,.
\eea
Thus $Z_{closed}=Z_{CS}$ implies that $\widehat{N}^{0}_{(0,1)}=2$ (consistent with \cite{KKV}) and 
\bea \label{relations}
&&\hspace{-2em}G_{1}(q,Q_{F})=Z_{1}(q,Q_{F})\,,\\ \nn
&&\hspace{-2em}G_{2}(q,Q_{F})=Z_{2}(q,Q_{F})- \frac{1}{2}Z_{1}(q,Q_{F})^{2}-\frac{1}{2}Z_{1}(q^2,Q_{F}^2)\,,\\ \nn
&&\hspace{-2em}G_{3}(q,Q_{F})=Z_{3}(q,Q_{F})+\frac{1}{3}Z_{1}(q,Q_{F})^3-Z_{1}(q,Q_{F})Z_{2}(q,Q_{F})-\frac{1}{3}Z_{1}(q^3,Q_{F}^3)\,,\\ \nn
&&\hspace{-2em}G_{4}(q,Q_{F})=Z_{4}(q,Q_{F})-\frac{1}{4}Z_{4}(q,Q_{F})^4+Z_{1}(q,Q_{F})^2Z_{2}(q,Q_{F})\\&&
\nn -Z_{1}(q,Q_{F})Z_{3}(q,Q_{F})-\frac{1}{2}Z_{1}(q,Q_{F})^2-\frac{1}{2}Z_{2}(q^2,Q_{F}^2)+\frac{1}{4}Z_{1}(q^2,Q_{F}^2)^2 \,. \\ \nn
\eea
The functions $Z_{k}(q,Q_{F})$ are easy to determine. From the above relations, we find that (for $k=1,2,3,4$)
\bea  \label{fgtp}
f^{(k)}_{g}(x)=\frac{P^{(k)}_{g}(x)}{(1-x)^{2g+4k-2}}\,,
\eea
where the functions $P^{(k)}_{g}(x)$ are finite at $x=1$. This behavior will become important when considering the field theory limit in the next subsection.

{\bf k=1:} In this case, since all curves $B+mF$ are of genus zero, it is
possible to obtain the invariants $N^{g}_{B+mF}$ directly. The moduli space of curves is
just $\IP^{2m+1}$ and therefore $N^{g}_{(1,m)}=-(2m+2)\delta_{g,0}$, hence we expect
\bea
f^{(1)}_{0}(x)=-\frac{2}{(1-x)^{2}}\,,\,\,\,\,\,f^{(1)}_{g>0}(x)=0\,.
\eea
This is exactly what we obtain from $Z_{1}(q,Q_{F})$.

{\bf k=2:}
In this case, by calculating $G_{2}(q,Q_{F})$, one can give an exact 
expression for all invariants,
\bea
f^{(2)}_{g}(x)=\frac{(3g+6)x^{g+1}+(6g+8)x^{g+2}+(3g+6)x^{g+3}}{(1-x)^{2g+6}(1+x)^{2}}\,.
\eea

{\bf k=3:} 
\bea
f^{(3)}_{g}(x)=\frac{x^{g+4}H^{(3)}_{g}(x)}{(1-x)^{2g+10}(1+x+x^{2})^{2}}\,,
\eea
where $H^{(3)}_{g}(x)$ is such that
\bea
H^{(3)}_{g}(x)=H^{(3)}_{g}(1/x)=\sum_{k=-g-3+[\frac{g+1}{2}]}^{g+3-[\frac{g+1}{2}]}A_{k}x^{k}\,.
\eea

\bea
H^{(3)}_{0}(x)&:=&8(x^3+x^{-3})+46(x^2+x^{-2})+100(x+x^{-1})+124\,,\\ \nn
H^{(3)}_{1}(x)&:=&68(x^3+x^{-3})+336(x^2+x^{-2})+692(x+x^{-1})+880\,\\ \nn
H^{(3)}_{2}(x)&:=&12(x^{4}+x^{-4})+436(x^{3}+x^{-3})+1874(x^{2}+x^{-2})\\\nn
& &+3736(x+x^{-1})+4732\,,\\ \nn
H^{(3)}_{3}(x)&:=&156(x^{4}+x^{-4})+2496(x^{3}+x^{-3})+9515(x^{2}+x^{-2})\\\nn
& &+18464(x+x^{-1})+23120\,,\\ \nn
H^{(3)}_{4}(x)&:=&16(x^{5}+x^{-5})+1304(x^{4}+x^{-4})+13368(x^{3}+x^{-3})\\\nn
&& +46118(x^{2}+x^{-2})+87180(x+x^{-1})+107852\,,\\ \nn
H^{(3)}_{5}(x)&:=&276(x^{5}+x^{-5})+8920(x^{4}+x^{-4})+68388(x^{3}+x^{-3})\\ \nn
&&+217040(x^{2}+x^{-2})+399888(x+x^{-1})+489312\,,\\ \nn
\label{k=3}
\eea

{\bf k=4:}

\bea
f^{(4)}_{g}(x)&=&\frac{x^{2g+6}H^{(4)}_{g}(x)}{(1-x)^{2g+14}(1+x)^{2g+6}}\,.\\
H^{(4)}_{0}(x)&=& 10(x^{5}+x^{-5}) + 208(x^{4}+x^{-4}) + 
1472(x^{3}+x^{-3})\\ \nn
&& + 5072(x^{2}+x^{-2}) + 10310(x+x^{-1})+12864\,,\\ \nn
H^{(4)}_{1}(x)&=& 300(x^{6}+x^{-6}) + 5392(x^{5}+x^{-5}) + 38977(x^{4}+x^{-4}) \\ \nn
&& +156500(x^{3}+x^{-3})+ 397376(x^{2}+x^{-2})\\ \nn
&& + 681628(x+x^{-1})+812710\,,\\ \nn
H^{(4)}_{2}(x)&=& 116(x^{8}+x^{-8}) + 7114(x^{7}+x^{-7}) + 105688(x^{6}+x^{-6})\\\nn
&& + 768492(x^{5}+x^{-5}) + 3394424(x^{4}+x^{-4})\\ \nn
&& + 10082352(x^{3}+x^{-3}) + 21285960(x^{2}+x^{-2}) \\ \nn
&&+ 32970906(x+x^{-1})+38079720\,, \\ \nn
H^{(4)}_{3}(x)&=& 15(x^{10}+x^{-10}) + 4560(x^{9}+x^{-9}) + 146856(x^{8}+x^{-8})\\\nn
&& + 1891720(x^{7}+x^{-7}) + 13702561(x^{6}+x^{-6})\\\nn
&& + 64651284(x^{5}+x^{-5}) + 214971644(x^{4}+x^{-4}) \\\nn
&&+ 527911700(x^{3}+x^{-3}) +985697328(x^{2}+x^{-2})\\\nn
&& + 1424513408(x+x^{-1})+1608879864\,,
\eea \bea
H^{(4)}_{4}(x)&=&1560(x^{11}+x^{-11}) + 120984(x^{10}+x^{-10})+ 2793760(x^{9}+x^{-9})\nn\\ \nn
&&+32488976(x^{8}+x^{-8}) + 233788052(x^{7}+x^{-7})\\ \nn
&& + 1156821600(x^{6}+x^{-6})+ 4187974036(x^{5}+x^{-5})\\ \nn
&& + 11538312784(x^{4}+x^{-4}) + 24830267172(x^{3}+x^{-3})\\\nn
&&+42464519560(x^{2}+x^{-2}) + 58354404732(x+x^{-1})\\
&&+64833791552\,,\\ \nn
H^{(4)}_{5}(x)&=& 276(x^{13}+x^{-13}) + 62765(x^{12}+x^{-12})+ 2707868(x^{11}+x^{-11})\\\nn
&& + 50597066(x^{10}+x^{-10}) + 
    545520996(x^{9}+x^{-9})\\ \nn
&& + 3898919969(x^{8}+x^{-8}) + 19995586316(x^{7}+x^{-7})\\ \nn
&& + 
    77346747002(x^{6}+x^{-6})+ 233315291868(x^{5}+x^{-5})\\ \nn
&& + 561626870823(x^{4}+x^{-4}) + 
    1096392376436(x^{3}+x^{-3})\\ \nn
&& + 1755278206204(x^{2}+x^{-2}) + 2321514065296(x+x^{-1})\\ \nn
&& + 2547127635094\,.\\ \nn
\label{k=4}
\eea

\subsection{Instantons}
In the last subsection, we saw that the generating functions for Gopakumar-Vafa invariants counting curves $kB+lF$ with fixed $k$ can conveniently be extracted from the open 
string partition function. As pointed out above, the world-sheet instantons wrapping these curves contribute to the k gauge instanton correction to the prepotential of the ${\cal N}=2$ theory. 

From the expansion of the relevant part of the topological string amplitude 
\bea
F_{instanton}(Q_{B},Q_{F},g_{s})=\sum_{(k,m)\neq (0,0)}\sum_{g=0}^{\infty}\sum_{n=0}^{\infty}\frac{\widehat{N}^{g}_{(k,m)}\,q^{n(1-g)}}{n(q^{n}-1)^{2-2g}}\,Q_{B}^{nk}Q_{F}^{nm}\,,
\eea
and recalling the field theory limit (\ref{fieldtheorylimit}) as $\beta \rightarrow 0$, we see that the divergence of $f_g^{(k)}(Q_F^n) \sim \beta^{2-2g-4k}$ (see equations (\ref{gfgv}), (\ref{fgtp})) is exactly cancelled by the $\beta$ dependence of $\frac{Q_{B}^{nk}}{(q^n-1)^{2-2g}} \sim \beta^{4kn+2g-2}$ for the case of single wrappings, $n=1$. Multiwrapping contributions vanish in the field theory limit.

The $k$-instanton contribution is thus given by ${\cal F}_k$,
\bea
{\cal F}_{k}&=&\lim_{\beta\rightarrow 0}
(\frac{\beta \Lambda}{2})^{4k}
\{\sum_{g=0}^{\infty}
\frac{f^{(k)}_{g}(1-2a\beta)}{(\beta \hbar)^{2-2g}}\}\,,\\ \nn
&=&a^{2}\,(\frac{\Lambda}{a})^{4k}\,c_{k}\,,
\eea
where
\bea  \label{cfromP}
c_{k}(\hbar,a)=\sum_{g=0}^{\infty}\frac{1}{\hbar^{2-2g}}\frac{P^{(k)}_{g}(1)}{2^{2g-2+8k}\,a^{2g}}\,.
\eea
The coefficient of $\hbar^{-2}$ in the expansion of $c_{k}(\hbar,a)$ is the $k$-instanton contribution to the prepotential. The coefficient of $\hbar^{0}$ is the $k$-instanton contribution to the coefficient of the $\int \mbox{Tr}R_{+}^{2}$ term in the effective action arising when the field theory is coupled to gravity. For the ${\cal N}=2$ D=4 SU(2) theory this has been
confirmed by comparing the results from the topologically twisted theory and matrix model calculations \cite{KMT,DST}. The field theory interpretation of the coefficients of $\hbar^{2g-2}$ ($g>1$) in the expansion of $c_{k}(\hbar,a)$ are as yet unclear (recall that these stem from the higher genus topological string amplitudes; in the low energy limit of type IIA, they describe the coupling of the graviphoton to $R_+^2$).

From the results of the previous subsection, we can easily calculate the first few instanton contributions.

Since $f^{(1)}_{g}(x)=\delta_{g,0}\,\frac{2}{(1-x)^{2}}$, we get
\bea
c_{1}(\hbar,a)=\frac{1}{\hbar^{2}}\,\frac{1}{2^{5}}\,.
\eea

For $c_{2}(\hbar,a)$, we extract $P^{(2)}_{g}(1)=3g+5$ from  
\bea
f^{(2)}_{g}(x)=\frac{(3g+6)x^{g+1}+(6g+8)x^{g+2}+(3g+6)x^{g+3}}{(1-x)^{2g+6}(1+x)^{2}} \,.
\eea
Thus we get
\bea
c_{2}(\hbar,a)&=&\nn\sum_{g=0}^{\infty}\frac{1}{\hbar^{2-2g}}\frac{3g+5}{2^{2g+14}a^{2g}}\\&=&\frac{1}{\hbar^{2}}\frac{5}{2^{14}}+\frac{1}{2^{13}a^{2}}+\hbar^{2}\,\frac{11}{2^{18}a^{4}}+\hbar^{4}\,\frac{7}{2^{19}a^{6}}+\cdots\,.
\eea

For $c_{3}(\hbar,a)$ we have, from equation (\ref{k=3}), the following expansion up to $g=10$,

\bea
c_{3}(\hbar,a)&=&\nn\sum_{g=0}^{\infty}\frac{1}{\hbar^{2-2g}}\frac{P^{(3)}_{g}(1)}{2^{2g+22}a^{2g}}\\\nn&=&\frac{1}{\hbar^{2}}\,\frac{3}{2^{18}}+\frac{1}{3\times2^{14}a^{2}}+\frac{\hbar^{2}}{a^{4}}\frac{117}{2^{22}}+\frac{\hbar^{4}}{a^{6}}\frac{293}{2^{23}}+\frac{\hbar^{6}}{a^{8}}\frac{8413}{3\times2^{26}}\\ \nn
&&+\frac{\hbar^{8}}{a^{10}}\frac{3261}{2^{26}}+\frac{\hbar^{10}}{a^{12}}\frac{59465}{2^{30}}+\frac{\hbar^{12}}{a^{14}}\frac{400493}{3\times2^{31}}+\frac{\hbar^{14}}{a^{16}}\frac{1184499}{2^{34}}\\ 
&&+\frac{\hbar^{16}}{a^{18}}\frac{650505}{2^{33}}+\frac{\hbar^{18}}{a^{20}}\frac{68040919}{3\times2^{38}}+\cdots\,.
\eea

For $c_{4}(\hbar,a)$ we have, from equation (\ref{k=4}), the following expansion up to $g=10$,
\bea
c_{4}(\hbar,a)&=&\sum_{g=0}^{\infty}\frac{1}{\hbar^{2-2g}}\frac{P^{(4)}_{g}(1)}{2^{2g+30}a^{2g}}\nn\\\nn&=&\frac{1}{\hbar^{2}}\,\frac{1469}{2^{31}}+\frac{1647}{2^{29}a^{2}}+\frac{\hbar^{2}}{a^{4}}\frac{171201}{2^{34}}+\frac{\hbar^{4}}{a^{6}}\frac{985823}{2^{35}}\\ \nn 
&&+\frac{\hbar^{6}}{a^{8}}\frac{42777927}{2^{39}}
+\frac{\hbar^{8}}{a^{10}}\frac{112053387}{2^{39}}+\frac{\hbar^{10}}{a^{12}}\frac{1147794293}{2^{41}}
\\ \nn
&&+\frac{\hbar^{12}}{a^{14}}\frac{5785079481}{2^{42}}+\frac{\hbar^{14}}{a^{16}}\frac{460910273265}{2^{47}}\\ 
&&+\frac{\hbar^{16}}{a^{18}}\frac{568311318115}{2^{46}}+\frac{\hbar^{18}}{a^{20}}\frac{22248943631667}{2^{50}}+\cdots.
\eea

\subsection{Field theory limit of Chern-Simons partition function} \label{ftl}
In the last subsection, we expressed the gauge instanton contributions in terms of curve counting formulas (equation (\ref{cfromP})), i.e. with the interpretation of the field theory as the low energy limit of type IIA in mind. We can equally well express the 
complete instanton partition function by taking the
field theory limit of $Z^{(m)}_{CS}(Q_{B},Q_{F};q)$ in (\ref{cspf}) directly, i.e. without the intermediate transcription to closed string quantities. In this spirit, it would be interesting to formulate a direct relationship between the three dimensional CS and the four dimensional ${\cal N}=2$ theory.

Unlike the previous subsection, we will here consider all three cases $\IF_m$, $m=1,2,3$ simultaneously. It will turn out that all $m$ dependence cancels in the field theory limit, as expected, {\it provided} we introduce a sign factor when relating the field theory to the geometric parameters, in the following way
\bea
Q_{B}=(-1)^{m}(\frac{\beta\Lambda}{2})^{4}\,,\,\,Q_{F}=e^{-2a\beta}\,,\,\,q=e^{-\beta\hbar}\,.  \label{mdep}
\eea
It would be interesting to justify the factor of $(-1)^m$ intrinsically.

Using these relations in equations (\ref{krr}) and (\ref{wl}),
\bea
\frac{K_{R_{1}R_{2}}(Q_{F})}{K_{\cdot\,\cdot}(Q_{F})}&=&\frac{{\cal L}_{R_{1}}(q)}{(q-1)^{l_{R_{1}}}}\frac{{\cal L}_{R_{2}}(q)}{(q-1)^{l_{R_{2}}}}\prod_{k}(1-q^{k}Q_{F})^{-C_{k}(R_{1},R_{2})}\,,
\eea
we obtain in the limit $\beta \rightarrow 0$
\bea 
\hspace{-2em}\frac{K_{R_{1}R_{2}}(Q_{F})}{K_{\cdot\,\cdot}(Q_{F})}\Eqn{=}\frac{1}{(\beta\hbar)^{l_{R_1}+l_{R_2}}}\{{\cal
  L}_{R_{1}}(1){\cal L}_{R_{2}}(1)\prod_{k}(2a\beta+\beta
\,k\,\hbar)^{-C_{k}(R_{1},R_{2})}+{\cal O}(\beta)\}\,,\nn \\ \hspace{-2em}
\Eqn{=}\frac{1}{\beta^{2l_{R_1}+2l_{R_2}}}\{\frac{{\cal L}_{R_{1}}(1){\cal L}_{R_{2}}(1)}{\hbar^{l_{R_{1}}+l_{R_{2}}}}\prod_{k}(2a+k\hbar)^{-C_{k}(R_{1},R_{2})}+{\cal O}(\beta)\}\,,
\eea
where we have used the fact that $\sum_{k}C_{k}(R_{1},R_{2})=l_{R_{1}}+l_{R_{2}}$. Thus,
\bea
{\cal Z}_{SU(2)}(\Lambda,a;\hbar)&:=&\nn\lim_{\beta\rightarrow 0}\frac{Z^{(m)}_{CS}(Q_{B},Q_{F};q)}{K_{\cdot \cdot}(Q_{F})^2}
\\& =&\sum_{R_{1},R_{2}}(\frac{\Lambda}{2\sqrt{\hbar}})^{4l_{1}+4l_{2}}\frac{{\cal L}_{R_{1}}(1)^{2}{\cal L}_{R_{2}}(1)^{2}}{\prod_{k}(2a+k\hbar)^{2C_{k}(R_{1},R_{2})}}\,,
\label{ourpf}
\eea
where the $m$-dependence cancels as promised. Hence, from equation (\ref{mdep}) it follows that all local $\IF_{m}$ 
3-folds yield the same results in the field theory limit for all genus.

\subsection{Relation with Nekrasov's conjecture}
The above instanton partition function agrees with
the partition function proposed by Nekrasov 
\cite{Nekrasov} and recently calculated in \cite{BFMT}.
To see this, recall that $Z^{(m)}_{CS}(T_B,T_F,q)$ is given by
\bea
\frac{Z^{(m)}_{CS}(T_B,T_F;q)}{K_{\cdot
    \cdot}(Q_{F})^{2}}&=&\sum_{R_{1},R_{2}}Q_{B}^{l_{R_{1}}+l_{R_{2}}}Q_{F}^{ml_{R_{2}}}(-1)^{m(l_{R_{1}}+l_{R_{2}})}\nn\\ & &\hspace{2em}q^{-\frac{m}{2}(\kappa_{R_{2}}+\kappa_{R_{1}})}
\frac{{\cal W}_{R_{1}}^{2}{\cal W}_{R_{2}^{T}}^{2}}{\prod_{k}(1-q^{k}Q_{F})^{2C_{k}(R_{1},R_{2}^{T})}}\,.
\eea
We have used $\kappa_{R^T}=-\kappa_R$.
The term $K_{\cdot \cdot}(Q)^{2}$ gives the 1-loop contribution to the prepotential in the field theory limit. We have divided the Chern-Simons partition function by $K_{\cdot \cdot}(Q)^{2}$ so that we only get the instanton contribution in the field theory limit. Now using the definition of ${\cal W}_{R}(q)$ given in equation (\ref{weq}) and the following identity
\bea 
\prod_{\infty>j>i\geq1}\frac{[\mu_{i}-\mu_{j}+j-i]}{[j-i]}&=&\prod_{d(\mu)\geq j>i\geq 1}\frac{[\mu_{i}-\mu_{j}+j-i]}{[j-i]}\,\,\prod_{i=1}^{d(\mu)}\prod_{\nu=1}^{\mu_{i}}\frac{1}{[\nu-i+d(\mu)]}\,,\nn\\\label{identity1}
\eea
where $\mu_i=0$ for $i > d(\mu)$, we see that for $q=e^{-\beta\hbar}$
\bea
{\cal W}_{R}^{2}(q)&=&2^{-2l_{R}}\,q^{\kappa_{R}/2}\prod_{i,j=1}^{\infty}\frac{\sinh\frac{\beta\hbar}{2}(\mu_{i}-\mu_{j}+j-i)}{\sinh\frac{\beta\hbar}{2}(j-i)} \,.
\eea
Note that care is required in treating the infinite product in order to obtain the factor $2^{-2l_R}$ correctly (e.g. by recourse to the RHS of equation (\ref{identity1})).
One can also check that the following identity holds, though we have not yet been able to derive it algebraically,
\bea \label{identity2}&&
\prod_{k}(1-q^{k}Q_{F})^{-2C_{k}(R_{1},R_{2}^{T})}=Q_{F}^{-l_{R_{1}}-l_{R_{2}}}2^{-2(l_{R_{1}}+l_{R_{2}})}q^{-\frac{1}{2}(\kappa_{R_{1}}-\kappa_{R_{2}})}\\ 
&&\prod_{l\neq n,i,j}\frac{\sinh\frac{\beta}{2}(a_{ln}+\hbar(\mu_{l,i}-\mu_{n,j}+j-i))}{\sinh\frac{\beta}{2}(a_{ln}+\hbar(j-i))}\,,\,\,l,n=1,2;\,i,j\geq 1\,.\nn
\eea
Again, the infinite product is to be interpreted along the lines of equation (\ref{identity1}). We have set $Q_{F}=e^{-2a\beta}$ and $a_{12}=-a_{21}=2a$.
Putting this all together, we obtain
\bea\nn&&
\frac{Z^{(m)}_{CS}(T_B,T_F=2a\beta;q=e^{-\beta\hbar})}{K_{\cdot \cdot}(Q_{F})^{2}}  =
\sum_{R_{1,2}}((-1)^{m}\frac{Q_{B}}{2^{4}Q_{F}})^{l_{R_{1}}+l_{R_{2}}}
Q_{F}^{m\,l_{R_{2}}}\\ \nn
&&\hspace{2em}q^{-\frac{m}{2}(\kappa_{R_{1}}+\kappa_{R_{2}})}\prod_{l,n=1,2}\prod_{i,j=1}^{\infty}\frac{\sinh\frac{\beta}{2}(a_{ln}+\hbar(\mu_{l,i}-\mu_{n,j}+j-i))}{\sinh\frac{\beta}{2}(a_{ln}+\hbar(j-i))}\,.
\eea
For $m=0$ (local $\IF_{0}$) this yields
\bea
Z^{(0)}_{CS}=K_{\cdot \cdot}(Q_{F})^{2}\,\sum_{R_{1},R_{2}}\varphi^{l_{R_{1}}+l_{R_{2}}}\,\prod_{l,n=1,2}\prod_{i,j=1}^{\infty}\frac{\sinh\frac{\beta}{2}(a_{ln}+\hbar(\mu_{l,i}-\mu_{n,j}+j-i))}{\sinh\frac{\beta}{2}(a_{ln}+\hbar(j-i))}\,,
\eea
where 
\bea
\varphi=\frac{Q_{B}}{2^{4}Q_{F}}=\frac{Q_{B}\,e^{2a\beta}}{2^{4}}\,.
\eea
This is exactly the form of the partition function derived by Nekrasov from an index 
calculation in \cite{Nekrasov} \footnote{A generalization of this to $SU(N)$ was also given by Nekrasov \cite{Nekrasov}
\bea
Z_{SU(N)}(\varphi,\beta)=\sum_{R_{1,\cdots,N}}\varphi^{l_{1}+\cdots +l_{N}}
\prod_{l,n=1}^{N}\prod_{i,j=1}^{\infty}\frac{\sinh\frac{\beta}{2}(a_{ln}+\hbar(\mu_{l,i}-\mu_{n,j}+j-i))}{\sinh
\frac{\beta}{2}(a_{ln}+\hbar(j-i))}\,.
\eea }.
In the field theory limit $Q_{B}=(\frac{\beta\Lambda}{2})^{4}$ with 
$\beta\rightarrow 0$, we get
\bea
{\cal Z}_{SU(2)}(\Lambda,a;\hbar)\Eqn{=}\mbox{lim}_{\beta\rightarrow 0}\frac{Z^{(m)}_{CS}(T_B=4\log(\frac{\beta\Lambda}{2}),T_F=2a\beta;q=e^{-\beta\hbar})}{K_{\cdot \cdot}(Q_{F})^2}\\ \nn
\Eqn{=}\sum_{R_{1},R_{2}}(\frac{\Lambda}{2})^{4(l_{R_{1}}+l_{R_{2}})}\prod_{l,n=1,2}\prod_{i,j=1}^{\infty}
\frac{a_{ln}+\hbar(\mu_{l,i}-\mu_{n,j}+j-i)}{a_{ln}+\hbar(j-i)}\,.
\eea
The fact that the expression is equal to the one 
given in equation (\ref{ourpf}) follows from the $\beta\rightarrow 0$ limit
of the identities given in equation (\ref{identity1}) and equation (\ref{identity2}).

The above partition function is a limit 
($\hbar_{1}=\hbar_{2}=\hbar$) of a more 
general partition function. 
 On the CS side 
(open string side) there exists a natural way of defining
a more general open string partition function by 
taking the coupling constant associated with the CS-theory
on each 3-cycle to be different. This more general partition function is given by
\bea
Z_{CS}(q_{1},q_{2},Q_{B},Q_{F})=\sum_{R_{1,2}}Q_{B}^{l_{1}+l_{2}}K_{R_{1}R_{2}}(q_{1},q_{2},Q_{F})^{2}\,,
\eea
where
\bea \label{k12}
K_{R_{1}R_{2}}(q_{1},q_{2},Q_{F})=\sum_{R}Q_{F}^{l_{R}}{\cal W}_{R_{1}R}(q_{1}){\cal W}_{RR_{2}}(q_{2}).
\eea
For $q_{1}=q_{2}$ we get back the original partition function.
Again we make an ansatz
\bea \label{k12f}
K_{R_{1}R_{2}}(q_{1},q_{2},Q_{F})={\cal W}_{R_{1}}(q_{1}){\cal W}_{R_{2}}(q_{2})
\mbox{Exp}\{f_{R_{1}R_{2}}^{n}(q_{1},q_{2})Q_{F}^{n}\}\,,
\eea
where
\bea f_{R_{1}R_{2}}^{n}(q_{1},q_{2})=
\frac{f_{R_{1}R_{2}}(q_{1}^{n},q_{2}^{n})}{n}\,.
\eea
Equating the coefficients of $Q_{F}$ in equations (\ref{k12}) and (\ref{k12f}) then yields
\bea\nn
f_{R_{1}R_{2}}(q_{1},q_{2})&=&\frac{{\cal W}_{R_{1},\tableau{1}}(q_{1})}{{\cal W}_{R_{1}}(q_{1})}
\frac{{\cal W}_{\tableau{1},R_{2}}(q_{2})}{{\cal W}_{R_{2}}(q_{2})}-{\cal W}_{\tableau{1}}(q_{1}){\cal W}_{\tableau{1}}(q_{2})\,\\\nn
&=&\frac{\sqrt{q_{1}q_{2}}}{(q_{1}-1)(q_{2}-1)}\{1+(q_{1}-1)\sum_{j=1}^{d_{1}}(q_{1}^{\mu_{j}-j}-q_{1}^{-j})\}\\
& &\{1+(q_{2}-1)\sum_{j=1}^{d_{2}}(q_{2}^{\nu_{j}-j}-q_{2}^{-j})\}  -\frac{\sqrt{q_{1}q_{2}}}{(q_{1}-1)(q_{2}-1)} \,.
\eea
The above expression can be simplified to the following form
\bea\nn
f_{R_{1}R_{2}}(q_{1},q_{2})&=&\sqrt{\frac{q_{2}}{q_{1}}}\frac{q_{1}-1}{q_{2}-1}f_{R_{1}}(q_{1})+\sqrt{\frac{q_{1}}{q_{2}}}\frac{q_{2}-1}{q_{1}-1}f_{R_{2}}(q_{2})\\
\nn && 
+\frac{(q_{1}-1)(q_{2}-1)}{\sqrt{q_{1}q_{2}}}f_{R_{1}}(q_{1})f_{R_{2}}(q_{2})\\ \nn
&=&\sum_{k_{1},k_{2}\in \IZ +\frac{1}{2}}C_{k_{1},k_{2}}(R_{1},R_{2})q_{1}^{k_{1}}q_{2}^{k_{2}}\,,
\eea
where $f_{R_{1}}(q)=\sum_{j=1}^{d_{1}}q^{-(j-1)}(1+q+\cdots
q^{\mu_{j}-1})$.

Then this more general partition function is given by
\bea
Z_{CS}(q_{1},q_{2},Q_{B},Q_{F})=\sum_{R_{1,2}}Q_{B}^{l_{1}+l_{2}}\frac{{\cal W}^{2}_{R_{1}}(q_{1}){\cal W}^{2}_{R_{2}}(q_{2})}
{\prod_{k_1,k_2}(1-q_{1}^{k_{1}}q_{2}^{k_{2}}Q_{F})
^{2C_{k_{1},k_{2}}(R_{1},R_{2})}} \,.
\eea
In the field theory limit we get
\bea
{\cal Z}(\Lambda,a;\hbar_{1},\hbar_{2})&=&\sum_{R_{1,2}}(\frac{\Lambda}{2})^{4(l_{1}+l_{2})}
\frac{{\cal L}^{2}_{R_{1}}(1)\,{\cal
    L}^{2}_{R_{2}}(1)}{\hbar^{2l_{1}}\,\hbar^{2l_{2}}}\nn\\& &\hspace{1em}\prod_{k_{1},k_{2}}(2a+\hbar_{1}k_{1}+\hbar_{2}k_{2})^{-2C_{k_{1},k_{2}}(R_{1},R_{2})} \,,
\eea
where we have used $\sum_{k_1,k_2} C_{k_1 k_2}(R_1,R_2)=l_1 + l_2$.

\section*{Acknowledgments}
AI would like to thank Mina Aganagic, Marcos Mari\~no and Cumrun Vafa for valuable 
discussions. AK would like to thank Allan Adams, Sergei Gukov, and David Morrison for helpful conversations. The research of AI was supported by NSF award NSF-DMS/00-74329. The 
research of AK was supported by the Department of Energy under contract number DE-AC03-76SF00515.

\section*{Appendix}
\subsection*{Details for local $\IP_2$}
\psfrag{one}{$x_1=0$}
\psfrag{two}{$x_2=0$}
\psfrag{three}{$x_3=0$}
\psfrag{four}{$x_4=0$}
\psfrag{five}{$x_5=0$}
\psfrag{six}{$x_6=0$}
\psfrag{seven}{$x_7=0$}
\psfrag{sone}{$\sigma_1$}
\psfrag{stwo}{$\sigma_2$}
\psfrag{sthree}{$\sigma_3$}
\psfrag{sfour}{$\sigma_4$}
\psfrag{sfive}{$\sigma_5$}
\psfrag{ssix}{$\sigma_6$}
\psfrag{sseven}{$\sigma_7$}

\begin{figure}[h]
\begin{center}\epsfxsize=.9\textwidth\leavevmode\epsfbox{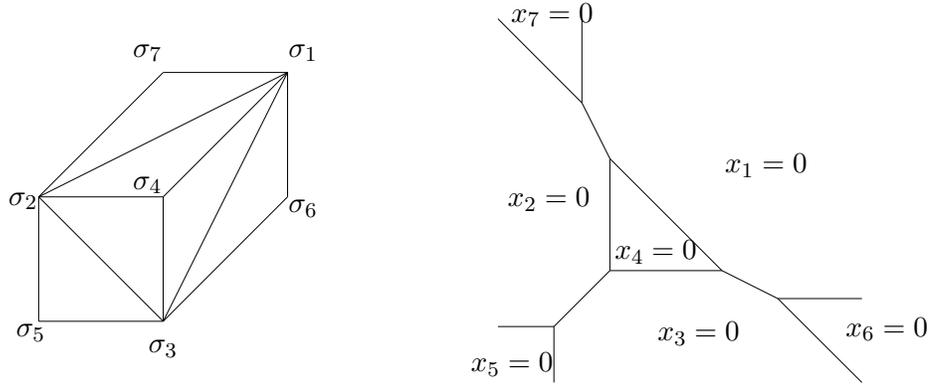}\end{center} 
\caption{\small The fan and web diagram for $\IB_3$. \label{b3}}
\end{figure}

From the toric diagram \figref{b3}, we can read off the following $\sigma$ model charges
\begin{eqnarray}
\begin{array}{rrrrrrrrr}
     ( & 1 & 1 & 1 & -3 & 0 & 0 & 0 & ) \\ \nn
     ( & 0 & -1 & -1 & 1 & 1 & 0 & 0 & ) \\ \nn
     ( & -1 & 0 & -1 & 1 & 0 & 1 & 0 & ) \\ \nn
     ( & -1 & -1 & 0 & 1 & 0 & 0 & 1 & ) \nn
\end{array} 
\end{eqnarray}
The monomials invariant under the four $U(1)$s encoded in these charges are generated by the following set
\begin{eqnarray}
\{x_1 x_2 x_3 x_4 x_5 x_6 x_7, \;x_1^2 x_2 x_4 x_6 x_7^2, \;x_1 x_2^2 x_4 x_5 x_7^2,\; x_2 x_3^2 x_4 x_5^2 x_6, \\ \nn
 x_2^2 x_3 x_4 x_5^2 x_7,\; x_1 x_3^2 x_4 x_5 x_6^2,\; x_1^2 x_3 x_4 x_6^2 x_7 \}
\end{eqnarray}
We cover the geometry with the 3 patches
\begin{eqnarray}
\mbox{Patch I:\;\;\;} x_3,x_5,x_6 \neq 0 \\ \nn
\mbox{Patch II:\;\;} x_1,x_6,x_7 \neq 0 \\ \nn
\mbox{Patch III:\;} x_2,x_5,x_7 \neq 0 \;.
\end{eqnarray}
In each patch, the invariant monomials are generated by the following set
\begin{eqnarray}
&&\mbox{Patch I:\hspace{1.5cm}} x= \frac{x_1 x_7}{x_3 x_5},\; y = x_2 x_3^2 x_4 x_5^2 x_6, \;u=\frac{x_2 x_7}{x_3 x_6},\; v=x_1 x_3^2 x_4 x_5 x_6^2 
\nn \\ &&\hspace{15em}\rightarrow \;\;xy =uv \\ \nn
&&\mbox{Patch II:\hspace{1.2cm}}\; \tx=\frac{x_3 x_5}{x_1 x_7}, \;\ty=x_1^2 x_2 x_4 x_6 x_7^2,\; \tu= \frac{x_2 x_5}{x_1 x_6},\; \tv = x_1^2 x_3 x_4 x_6^2 x_7 
\\\nn&&\hspace{15em}\rightarrow\;\; \tx \ty = \tu \tv \\ \nn
&&\mbox{Patch III:\hspace{0.5cm}} x'=\frac{x_3 x_6}{x_2 x_7},\; y'=x_1 x_2^2 x_4 x_5 x_7^2, \;u' = \frac{x_1 x_6}{x_2 x_5}, \;v' = x_2^2 x_3 x_4 x_5^2 x_7 
\\\nn&&\hspace{15em}\rightarrow\;\; x' y' = u' v'\,.
\end{eqnarray}
The transition functions in the overlap of the patches can easily be read off from these equations:
\begin{align}
x&=\frac{1}{\tx}\;, \qquad y = \tx \tu \tv \;, \qquad u =\frac{\ty}{\tv}\;, \qquad v= \tx \tv  \;,  \\\nn
\tx &= \frac{x'}{u'}\;, \qquad \ty = y' u'\;\qquad  \tu = \frac{1}{u'}\;,\qquad \tv = {u'}^2 v' \;,\\ \nn
x' &= \frac{v}{xy}\,,\qquad y'= u x y\;, \qquad u'=\frac{x}{u} \;, \qquad v'= \frac{x y^2}{v}\;.
\end{align}
Note that the relation of the phases of the coordinates in different patches can be read off from the web diagram.
We now perform the following deformations on the constraint equations
\begin{eqnarray}
xy &=& uv + \mu_1 \,,\\
\tx \ty  &=& \tu \tv + \mu_2 \,,\\
x' y' + \mu_1 &=& u' v' + \mu_2 \,. 
\end{eqnarray}
To enforce the relation $z=xy=\tx \ty= x'y' + \mu_1$, we must also deform the transition functions,
\begin{align}
x&=\frac{1}{\tx}\;,\qquad y= \tx \tu \tv + \mu_2 \tx\;,\qquad u =\frac{\ty}{\tv}-\frac{\mu_1}{\tx \tv}\;,\qquad v= \tx \tv \;,\\ \nn
\tx &= \frac{x'}{u'}\;,\qquad \ty = y' u'+ \mu_1 \frac{u'}{x'} \;, \qquad\tu = \frac{1}{u'}\;,\qquad \tv = {u'}^2 v' \;,\\ \nn
x' &= \frac{v}{x}\frac{1}{y-\frac{\mu_2}{x}}\;,\;\;\; y'= u x (y-\frac{\mu_2}{x})\;, \;\;\; u'=\frac{v}{y-\frac{y-\mu_2}{x}}\;,\;\;\; v'= \frac{x}{v}(y-\frac{\mu_2}{x})^2 \;.
\end{align}
These deformations maintain the phase relations of the coordinates in the three patches.

\subsection*{local $\IF_{0}$}
Here we list functions $H^{(n)}_{g}(x)$ for $n=3,4$ and 
$g=0,\dots, 15$.
\bea \nn
H^{(3)}_{6}(x)&=&20(x^{6}+x^{-6})+2860(x^{5}+x^{-5})+54344(x^{4}+x^{-4})\\ \nn
&&+338204(x^{3}+x^{-3})+1000022(x^{2}+x^{-2})\\ \nn
&&+1797108(x+x^{-1})+2177844\,\\ \nn
H^{(3)}_{7}(x)&=&428(x^6+x^{-6})+23152(x^5+x^{-5})+306968(x^4+x^{-4})\\ \nn
&&
+1629392(x^3+x^{-3})+4532536(x^2+x^{-2})\\ \nn
&&+7953136(x+x^{-1})+9556104\,\\ \nn
H^{(3)}_{8}(x)&=&24(x^7+x^{-7})+5296(x^6+x^{-6})+161552(x^5+x^{-5})
\\ \nn
&&+1643952(x^4+x^{-4})+7688416(x^3+x^{-3})\\ \nn
&&+20272814(x^2+x^{-2})+
34777628(x+x^{-1})+41468492\,,\\ \nn
H^{(3)}_{9}(x)&=&612(x^7+x^{-7})+49736(x^6+x^{-6})+1020652(x^{5}+x^{-5})\\ \nn
&&+
8462384(x^4+x^{-4})+35667972(x^3+x^{-3})\\ \nn
&&+89686864(x^{2}+x^{-2})+
150623428(x+x^{-1})+178358464\,,\\ \nn
H^{(3)}_{10}(x)&=&28(x^8+x^{-8})+8804(x^7+x^{-7})+393728(x^6+x^{-6})\\ \nn
&&+6006028(x^5+x^{-5})+42249200(x^{4}+x^{-4})\\ \nn
&&+
163151268(x^{3}+x^{-3})+393139474(x^{2}+x^{-2})\\ \nn
&&+647251192(x+x^{-1})+761564668\,,\\ \nn
H^{(3)}_{11}(x)&=&828(x^{8}+x^{-8})+94240(x^{7}+x^{-7})+2772832(x^{6}+x^{-6})\\ \nn
&&
+33511552(x^{5}+x^{-5})+205862000(x^{4}+x^{-4})\\ \nn
&&+737434848(x^{3}+x^{-3})
+1709869228(x^{2}+x^{-2})\\ \nn
&&+2763195520(x+x^{-1})+3232117936\,,\\ \nn
H^{(3)}_{12}(x)&=&32(x^{9}+x^{-9})+13576(x^{8}+x^{-8})+836744(x^{7}+x^{-7})\\ \nn
&&
+17934000(x^{6}+x^{-6})+179430424(x^{5}+x^{-5})\\ \nn
&&+983373728(x^{4}+x^{-4})
+3299321448(x^{3}+x^{-3})\\ \nn
&&+7386742102(x^{2}+x^{-2})+11731393228(x+x^{-1})+13647278764\,,\\ \nn
\widehat{H}^{(3)}_{13}(x)&=&1076(x^{9}+x^{-9})+163256(x^{8}+x^{-8})+
6521908(x^{7}+x^{-7})\\ \nn
&&+108732192(x^{6}+x^{-6})+929691024(x^{5}+x^{-5})\\ \nn
&&+
4620614976(x^{4}+x^{-4})+14631592752(x^{3}+x^{-3})\\ \nn
&&+31725047776(x^{2}+x^{-2})
+49571743936(x+x^{-1})+57373130752\,,\\ \nn
H^{(3)}_{14}(x)&=&36(x^{10}+x^{-10})+19804(x^{9}+x^{-9})+1608728(x^{8}+x^{-8})
\\ \nn
&&+46170268(x^{7}+x^{-7})+626678580(x^{6}+x^{-6})\\ \nn
&&+4690016816(x^{5}+x^{-5})
+21410770496(x^{4}+x^{-4})\\\nn
&&+64389042000(x^{3}+x^{-3})+135558792610(x^{2}+x^{-2})\\ \nn
&&+208614265444(x+x^{-1})+240290143540\,,\eea \bea \nn
H^{(3)}_{15}(x)&=&1356(x^{10}+x^{-10}) + 264400(x^9+x^{-9}) + 13772840(x^8+x^{-8}) \\ \nn
&& + 303539824(x^7+x^{-7})+ 3467940668(x^6+x^{-6})  \\ \nn
&&+ 23141374272(x^5+x^{-5}) + 
    98036368128(x^4+x^{-4})\\ \nn
&& + 281441631392(x^3+x^{-3}) + 576623540332(x^2+x^{-2})\\ \nn
&& + 
    874793870512(x+x^{-1})\\ \nn
&& +1003092794248\,.
\\ \nn
H^{(4)}_{6}(x)&=& 20(x^{15}+x^{-15})+ 20776(x^{14}+x^{-14}) + 1802422(x^{13}+x^{-13})\\ \nn
&& + 55118752(x^{12}+x^{-12})+ 887791786(x^{11}+x^{-11}) \\ \nn
&&+ 9032681160(x^{10}+x^{-10}) + 64154932056(x^{9}+x^{-9})\\ \nn
&& + 
    338318599184(x^{8}+x^{-8})+ 1379961590592(x^{7}+x^{-7})\\ \nn
&& + 4479333446968(x^{6}+x^{-6}) + 
    11809327834558(x^{5}+x^{-5}) \\ \nn
&&+ 25665330292512(x^{4}+x^{-4}) + 46479319690946(x^{3}+x^{-3})\\ \nn
&&+ 
    70678653611736(x^{2}+x^{-2}) + 90710040683380(x+x^{-1})\\ \nn
&&+  98545938094688\,,\\ \nn
H^{(4)}_{7}(x)&=& 4266(x^{16}+x^{-16}) + 839844(x^{15}+x^{-15}) + 43543660(x^{14}+x^{-14})\\ \nn
&& + 1056778816(x^{13}+x^{-13})+ 15241810336(x^{12}+x^{-12})\\ \nn
&& + 148139062508(x^{11}+x^{-11}) + 1046335719584(x^{10}+x^{-10})\\ \nn
&&+ 
    5642396903464(x^{9}+x^{-9}) + 24036080776384(x^{8}+x^{-8})\\ \nn
&& + 82894576962384(x^{7}+x^{-7}) +235675324886872(x^{6}+x^{-6})\\ \nn
&& + 559907882759836(x^{5}+x^{-5}) + 1122949968857408(x^{4}+x^{-4})\\ \nn
&& + 1915763227378616(x^{3}+x^{-3})+ 2795404761655196(x^{2}+x^{-2})\\ \nn
&& + 3501738224591460(x+x^{-1})+    3773935482414956\,,\\ \nn
H^{(4)}_{8}(x)&=&496(x^{18}+x^{-18}) + 273912(x^{17}+x^{-17}) + 25434960(x^{16}+x^{-16})\\ \nn
&& + 947885136(x^{15}+x^{-15})+ 
    19453650776(x^{14}+x^{-14})\\ \nn
&& + 257567625014(x^{13}+x^{-13}) + 2412441992944(x^{12}+x^{-12})\\ \nn
&& + 
    16956815818608(x^{11}+x^{-11}) + 93128479478936(x^{10}+x^{-10})\\ \nn
&&+ 
    411350743539242(x^{9}+x^{-9}) + 1492803031085792(x^{8}+x^{-8})\\ \nn
&& + 
    4523350413679026(x^{7}+x^{-7}) + 11586657433749704(x^{6}+x^{-6})\\ \nn
&& + 
    25330337013080364(x^{5}+x^{-5}) + 47609815386901776(x^{4}+x^{-4})\\ \nn
&&+ 
    77364284922244622(x^{3}+x^{-3}) + 109129776774805848(x^{2}+x^{-2})\\ \nn
&& + 
    134003186940618748(x+x^{-1})  +143470224614509472\,,\eea \bea \nn
H^{(4)}_{9}(x)&=&  25(x^{20}+x^{-20}) + 61316(x^{19}+x^{-19}) + 11068374(x^{18}+x^{-18})\\ \nn
&&\hspace{-3em}  + 652429176(x^{17}+x^{-17}) + 
    19259125692(x^{16}+x^{-16})\\ \nn&&\hspace{-3em}  + 347810000852(x^{15}+x^{-15})+ 4300723919714(x^{14}+x^{-14})\\ \nn
&&\hspace{-3em}  + 
    39070905330120(x^{13}+x^{-13}) + 273457920809930(x^{12}+x^{-12}) \\ \nn
&&\hspace{-3em} 
+ 
    1524969346570920(x^{11}+x^{-11}) + 6945674565624384(x^{10}+x^{-10})\\ \nn
&&\hspace{-3em} 
 + 
    263257650262280089(x^{9}+x^{-9}) + 84243018258047546(x^{8}+x^{-8})\\ \nn
&& \hspace{-3em} 
+ 
    230185454300541696(x^{7}+x^{-7}) + 541844957793011864(x^{6}+x^{-6})\\ \nn
&& \hspace{-3em} 
+ 
    1106543100329064720(x^{5}+x^{-5}) + 1971231278784742349(x^{4}+x^{-4})\\ \nn
&&\hspace{-3em} 
 + 
    3076166028299942844(x^{3}+x^{-3}) + 4218295082475696624(x^{2}+x^{-2})\\ \nn
&&\hspace{-3em} 
+ 
    5093897002657007052(x+x^{-1}) +5423687898801824948\,,\nn\\\nn
H_{10}(x)&=& 8982(x^{21}+x^{-21}) + 3578472(x^{20}+x^{-20}) + 348034206(x^{19}+x^{-19})\\ \nn
&&\hspace{-3em} 
+ 
    15037643104(x^{18}+x^{-18})+ 372747895080(x^{17}+x^{-17})\\ \nn
&&\hspace{-3em}  + 6084321194460(x^{16}+x^{-16}) 
+ 
    71140851380158(x^{15}+x^{-15}) \\ \nn
&&\hspace{-3em}  + 629960905795240(x^{14}+x^{-14})
+ 
    4392643781205418(x^{13}+x^{-13})\\ \nn
&& \hspace{-3em}  + 24815217549021088(x^{12}+x^{-12}) + 
    116039047539664182(x^{11}+x^{-11})\\ \nn
&& \hspace{-3em}  + 456636168407024952(x^{10}+x^{-10}) 
+ 
    1532000433014962822(x^{9}+x^{-9})\\ \nn
&&\hspace{-3em}  + 4427463940221909760(x^{8}+x^{-8})
 + 
    11113497261742849754(x^{7}+x^{-7})\\ \nn
&&\hspace{-3em}  + 24391020745301746344(x^{6}+x^{-6})
 + 47054462823805306748(x^{5}+x^{-5}) \\ \nn
&&\hspace{-3em}  + 80129270910314118520(x^{4}+x^{-4})
 + 
    120842120340657126960(x^{3}+x^{-3}) \\ \nn
&&\hspace{-3em}  + 161785295782709385240(x^{2}+x^{-2})
+ 
    192612749016106866042(x+x^{-1})\\ \nn
&&\hspace{-3em}  + 204120388510473361224\,,\\\nn
\hspace{-4em}H_{11}(x)&=&776(x^{23}+x^{-23}) + 847738(x^{22}+x^{-22}) + 144344832(x^{21}+x^{-21})\\ \nn
&&  \hspace{-3em} 
+ 9357879388(x^{20}+x^{-20}) + 
    321867505044(x^{19}+x^{-19})\\ \nn
&& \hspace{-3em}  + 6960563372879(x^{18}+x^{-18}) 
+ 104659625277760(x^{17}+x^{-17})\\ \nn
&& \hspace{-3em}  + 
    1167957443029360(x^{16}+x^{-16})
 + 10119596098395264(x^{15}+x^{-15}) \\ \nn
&&\hspace{-3em}  + 
    70330233541355881(x^{14}+x^{-14})
+ 401754102187227592(x^{13}+x^{-13})\\ \nn
&& \hspace{-3em}  + 
    1922054663797082448(x^{12}+x^{-12}) 
+ 7815436357972473172(x^{11}+x^{-11}) \\ \nn
&& \hspace{-3em} + 
    27328823525971791460(x^{10}+x^{-10}) 
+ 82962081227932553344(x^{9}+x^{-9})\\ \nn
&& \hspace{-3em}  + 
    220327585105407247080(x^{8}+x^{-8}) 
+ 515123507705963450920(x^{7}+x^{-7})\\ \nn
&& \hspace{-3em}  + 
    1065676633098908467490(x^{6}+x^{-6}) 
+ 1958880323008004536300(x^{5}+x^{-5})\\ \nn
&&\hspace{-3em}   + 
    3209930502325208372372(x^{4}+x^{-4})
+ 4701228070777034064356(x^{3}+x^{-3})\\ \nn
&&\hspace{-3em}   + 
    6165909219169186575432(x^{2}+x^{-2})
 + 7251680674388081571600(x+x^{-1})\\ \nn
&&\hspace{-3em}  +7653871498185950485200\,, \eea \bea \nn
\hspace{-4em}H_{12}(x)&=&30(x^{25}+x^{-25}) + 142952(x^{24}+x^{-24}) + 46397758(x^{23}+x^{-23})\\ \nn
&& \hspace{-4.5em} 
 + 4666992904(x^{22}+x^{-22}) + 
    226254790488(x^{21}+x^{-21})\\ \nn
&& \hspace{-4.5em}
 + 6528038807672(x^{20}+x^{-20}) + 126479209707000(x^{19}+x^{-19}) \\ \nn
&& \hspace{-4.5em}
+ 
    1776522953134224(x^{18}+x^{-18}) + 19057077119890680(x^{17}+x^{-17}) \\ \nn
&& \hspace{-4.5em}
+ 
    162049457830972496(x^{16}+x^{-16}) + 1122934839735652880(x^{15}+x^{-15})\\ \nn
&& \hspace{-4.5em}
 + 
    6476817384316875496(x^{14}+x^{-14}) + 31612278725628725440(x^{13}+x^{-13}) \\ \nn
&& \hspace{-4.5em}
 + 
    132300992609304637960(x^{12}+x^{-12}) + 479854874693771928080(x^{11}+x^{-11})\\ \nn
&& \hspace{-4.5em}
 + 
    1521505193447041212400(x^{10}+x^{-10}) + 4247761951357534695194(x^{9}+x^{-9})\\ \nn
&& \hspace{-4.5em}
 + 
    10503538092109971310456(x^{8}+x^{-8}) + 23116405003776992480010(x^{7}+x^{-7}) \\ \nn
&& \hspace{-4.5em}
+ 
    45463376413349088082344(x^{6}+x^{-6}) + 80165702837137187676152(x^{5}+x^{-5})\\ \nn
&& \hspace{-4.5em}
 + 
    127072760262836311167104(x^{4}+x^{-4})+ 181450624018663596285928(x^{3}+x^{-3})\\ \nn
&& \hspace{-4.5em}
 + 
    233770042020772015684072(x^{2}+x^{-2}) + 272031823542995558909304(x+x^{-1})\\ \nn
&&\hspace{-4.5em} + 286109526001880184530336\,,
\\\nn
\hspace{-4em}H_{13}(x)&=&16248(x^{26}+x^{-26}) + 11441484(x^{25}+x^{-25}) + 1867609602(x^{24}+x^{-24}) \\ \nn
&&\hspace{-4.5em} 
+ 
    130313168272(x^{23}+x^{-23}) + 5068617330184(x^{22}+x^{-22}) \\ \nn
&& \hspace{-4.5em}
+ 127122681176684(x^{21}+x^{-21}) + 
    2249626573376071(x^{20}+x^{-20})\\ \nn
&& \hspace{-4.5em}
 + 29833194044232700(x^{19}+x^{-19})  + 
    309352842642427198(x^{18}+x^{-18}) \\ \nn
&& \hspace{-4.5em}
+ 2587963124320382488(x^{17}+x^{-17}) + 
    17886698663434866553(x^{16}+x^{-16}) \\ \nn
&&\hspace{-4.5em} + 
 104041685392475379304(x^{15}+x^{-15})  + 
    516863496360020461678(x^{14}+x^{-14}) \\ \nn
&&\hspace{-4.5em} 
+ 2219175312759092765388(x^{13}+x^{-13}) + 
    8315148255251476062787(x^{12}+x^{-12})\\ \nn
&&\hspace{-4.5em} 
 + 27408849538963007252540(x^{11}+x^{-11}) + 
    80010468270453478393296(x^{10}+x^{-10})\\ \nn
&&\hspace{-4.5em} 
 + 207995282777817187829848(x^{9}+x^{-9}) + 
    483761555487399956706890(x^{8}+x^{-8}) \\ \nn
&&\hspace{-4.5em} 
+ 1010575072306854645122796(x^{7}+x^{-7}) + 
    1902256171458206034909248(x^{6}+x^{-6})\\ \nn
&&\hspace{-4.5em} 
 + 3235110563313776191505496(x^{5}+x^{-5})  + 
    4981575619223563155734966(x^{4}+x^{-4})\\ \nn
&&\hspace{-4.5em} 
 + 6957331918466546588154936(x^{3}+x^{-3}) + 
    8824248101674959733825220(x^{2}+x^{-2}) \\ \nn
&&\hspace{-4.5em} 
+ 10173350240915563250118496(x+x^{-1})      +        10666791272137599144340470\,
\\\nn
\hspace{-4em}H_{14}(x)&=&1116(x^{28}+x^{-28}) + 2123034(x^{27}+x^{-27}) + 597858776(x^{26}+x^{-26})\\ \nn
&&\hspace{-4.5em} 
 + 
    61694966866(x^{25}+x^{-25}) + 3280150367480(x^{24}+x^{-24})\\ \nn
&&\hspace{-4.5em}  + 107249898982190(x^{23}+x^{-23})
 + 
    2398456260768240(x^{22}+x^{-22})\\ \nn
&&\hspace{-4.5em}  + 39334297729267612(x^{21}+x^{-21}) 
+ 
    496590877057032204(x^{20}+x^{-20}) \\ \nn
&&\hspace{-4.5em} + 4999965296583002358(x^{19}+x^{-19})
 + 
    41233256060532653688(x^{18}+x^{-18})\\ \nn
&&\hspace{-4.5em} + 284316242947879915100(x^{17}+x^{-17}) 
 + 
 1666176504493064118336(x^{16}+x^{-16})\\ \nn
&&\hspace{-4.5em}  + 8408482528507356296876(x^{15}+x^{-15}) 
+ 
    36936675268551592344472(x^{14}+x^{-14})\\ \nn
&&\hspace{-4.5em}  + 142494162691476258768526(x^{13}+x^{-13})
 + 
    486350783413982018389556(x^{12}+x^{-12}) \\ \nn
&&\hspace{-4.5em} + 1477790555674438148532576(x^{11}+x^{-11}) 
+ 
    4018472087633195293071392(x^{10}+x^{-10})\\ \nn
&&\hspace{-4.5em}  + 9822352746135144088586634(x^{9}+x^{-9}) 
+ 
    21662099226298401054640968(x^{8}+x^{-8}) \eea 
\begin{small}
\bea\ \nn
&&\hspace{-4.5em} + 43239893498911642392701526(x^{7}+x^{-7})
 + 78327490163691445755024792(x^{6}+x^{-6})\\ \nn
&&\hspace{-4.5em}  + 129044537123542177935391914(x^{5}+x^{-5}) 
+ 
    193702544707394610401774948(x^{4}+x^{-4})\\ \nn
&&\hspace{-4.5em}  + 265286854476302190398619068(x^{3}+x^{-3})
 + 
    331854109645644252336639088(x^{2}+x^{-2}) \\ \nn
&&\hspace{-4.5em} + 379452019419503320388448152(x+x^{-1})
 +  396769857245681885554908928\,,
\\\nn
\hspace{-1em}H_{15}(x)&=&  35(x^{30}+x^{-30}) + 286744(x^{29}+x^{-29}) + 151909916(x^{28}+x^{-28})\\ \nn
&&\hspace{-4.5em}  
+ 24014981300(x^{27}+x^{-27}) + 
    1776678078197(x^{26}+x^{-26})\\ \nn
&&\hspace{-4.5em}  + 76462540562016(x^{25}+x^{-25})
 + 
    2171568812362050(x^{24}+x^{-24})\\ \nn
&&\hspace{-4.5em}  + 44128251927349080(x^{23}+x^{-23})
 + 
    678229429084167029(x^{22}+x^{-22})\\ \nn
&&\hspace{-4.5em}  + 82055061961991846889(x^{21}+x^{-21}) 
+ 
    80513486253657266932(x^{20}+x^{-20})\\ \nn
&&\hspace{-4.5em}  + 655602921413643536680(x^{19}+x^{-19})
 + 
    4511043260336323140811(x^{18}+x^{-18})\\ \nn
&&\hspace{-4.5em}  + 26612229682448017966692(x^{17}+x^{-17}) 
+ 
    136204891358039983634884(x^{16}+x^{-16})\\ \nn
&&\hspace{-4.5em}  + 610736662112601476054180(x^{15}+x^{-15}) 
+ 
    2418823210255342971140237(x^{14}+x^{-14}) \\ \nn
&&\hspace{-4.5em} + 8519606663089707541414760(x^{13}+x^{-13})
+ 
    26842281493690029147172220(x^{12}+x^{-12}) \\ \nn
&&\hspace{-4.5em} + 76023106991445943817131912(x^{11}+x^{-11}) 
+ 
    194367691907203562148213795(x^{10}+x^{-10})\\ \nn
&&\hspace{-4.5em}  + 
    450207515759262474725900584(x^{9}+x^{-9}) 
+ 947630554842568898139897502(x^{8}+x^{-8}) \\ \nn
&&\hspace{-4.5em} + 
    1817312857981270397216413376(x^{7}+x^{-7}) 
+ 
    3182248997707767165484154515(x^{6}+x^{-6})\\ \nn
&&\hspace{-4.5em}  
+
    5097338277427026638104847412(x^{5}+x^{-5}) 
 + 
    7480058119725431973984814452(x^{4}+x^{-4})\\ \nn
&&\hspace{-4.5em} 
 + 
    10067787731012537908374528440(x^{3}+x^{-3}) 
+ 
    12440072163092579373241505125(x^{2}+x^{-2})\\ \nn
&&\hspace{-4.5em} 
 + 14120405833090393940001592152(x+x^{-1})     
+14728907025122899597601666232\,.
\eea
\end{small}

\subsection*{local $\IF_{1}$} 

Here we list the functions $H^{(n)}_{g}(x)$ for $n=3$ and 
$g=0,\cdots 15$. The 
invariants of local $\IF_{0}$ and local $\IF_{1}$ are related to
each other for curves with even wrapping number on the 
base, as can be seen easily by using the affine $E_{8}$ Weyl symmetry of local $\frac{1}{2}$ K3. 
Thus the $n=4$ case for local $\IF_{1}$ can be derived from 
the functions $H^{(4)}_{g}(x)$ defined in the previous subsection.
\bea
H^{(3)}_{6}(x)&=&348(x^{11/2}+x^{-11/2}) + 14800(x^{9/2}+x^{-9/2})\\ \nn
&&  + 151712(x^{7/2}+x^{-7/2}) 
+626785(x^{5/2}+x^{-5/2}) \\ \nn
&& + 1412346(x^{3/2}+x^{-3/2}) +
2075489(x^{1/2}+x^{-1/2})\,,\\ \nn
H^{(3)}_{7}(x)&=& 22(x^{13/2}+x^{-13/2})+ 3956(x^{11/2}+x^{-11/2})+ 96574(x^{9/2}+x^{-9/2}) \\ \nn
&& + 778054(x^{7/2}+x^{-7/2}) 
+ 2906210(x^{5/2}+x^{-5/2}) \\ \nn
&& + 6311298(x^{3/2}+x^{-3/2}) + 9127550(x^{1/2}+x^{-1/2})\,,\eea \bea\ \nn
H^{(3)}_{8}(x)&=& 516(x^{13/2}+x^{-13/2}) + 34576(x^{11/2}+x^{-11/2}) \\ \nn
&& + 576860(x^{9/2}+x^{-9/2}) 
+ 3862860(x^{7/2}+x^{-7/2})\\ \nn
&& + 13270807(x^{5/2}+x^{-5/2}) + 27851098(x^{3/2}+x^{-3/2}) \\ \nn
&& 
+ 39687211(x^{1/2}+x^{-1/2})\,,\\\nn
H^{(3)}_{9}(x)&=&26(x^{15/2}+x^{-15/2}) + 6904(x^{13/2}+x^{-13/2})\\ \nn
&& + 257348(x^{11/2}+x^{-11/2}) 
+ 3236514(x^{9/2}+x^{-9/2}) \\ \nn
&& + 18699598(x^{7/2}+x^{-7/2})+ 59831120(x^{5/2}+x^{-5/2})\\ \nn
&& 
 + 121660400(x^{3/2}+x^{-3/2}) + 170998970(x^{1/2}+x^{-1/2})\,,\\\nn
H^{(3)}_{10}(x)&=&716(x^{15/2}+x^{-15/2})+ 69360(x^{13/2}+x^{-13/2}) \\ \nn
&& + 1717944(x^{11/2}+ x^{-11/2}) 
+  17331516(x^{9/2}+x^{-9/2}) \\ \nn
&& + 88704772(x^{7/2}+x^{-7/2})+ 266850161(x^{5/2}+x^{-5/2}) \\ \nn
&& 
+ 526995098(x^{3/2}+x^{-3/2})+ 731312489(x^{1/2}+x^{-1/2})\,,
\\ \nn
H^{(3)}_{11}(x)&=& 30(x^{17/2}+x^{-17/2}) + 11020(x^{15/2}+x^{-15/2})\\ \nn
&&  + 582314(x^{13/2}+x^{-13/2}) 
+ 10602230(x^{11/2}+x^{-11/2})\\ \nn
&&  + 89528346(x^{9/2}+x^{-9/2}) + 413833342(x^{7/2}+x^{-7/2})\\ \nn
&&  
+ 1179215768(x^{5/2}+x^{-5/2}) + 2266727282(x^{3/2}+x^{-3/2})\\ \nn
&&  + 3108299684(x^{1/2}+x^{-1/2})\,,\\\nn
H^{(3)}_{12}(x)&=&948(x^{17/2}+x^{-17/2}) + 125232(x^{15/2}+x^{-15/2})\\ \nn
&&  + 4318452(x^{13/2}+x^{-13/2}) 
+ 61670124(x^{11/2}+x^{-11/2}) \\ \nn
&& + 449427764(x^{9/2}+x^{-9/2}) + 
1903914844(x^{7/2}+x^{-7/2})\\ \nn
&& 
 + 5169507095(x^{5/2}+x^{-5/2}) + 9691341866(x^{3/2}+x^{-3/2})\\ \nn
&&  + 13142378339(x^{1/2}+x^{-1/2})\,,\\\nn
H^{(3)}_{13}(x)&=&34(x^{19/2}+x^{-19/2}) + 16496(x^{17/2}+x^{-17/2})\\ \nn
&& + 1172832(x^{15/2}+x^{-15/2}) 
+ 29239874(x^{13/2}+x^{-13/2})\\ \nn
&&  + 342547768(x^{11/2}+x^{-11/2})+ 
2204219504(x^{9/2}+x^{-9/2}) \\ \nn
&& 
+ 8656118128(x^{7/2}+x^{-7/2}) +
 22505083592(x^{5/2}+x^{-5/2})\\ \nn
&&  + 41221499060(x^{3/2}+x^{-3/2}) 
+ 55320776984(x^{1/2}+x^{-1/2})\,,\\\nn
H^{(3)}_{14}(x)&=&1212(x^{19/2}+x^{-19/2}) + 209296(x^{17/2}+x^{-17/2})\\ \nn
&& + 9589584(x^{15/2}+x^{-15/2}) 
+ 184678652(x^{13/2}+x^{-13/2}) \\ \nn
&& + 1833812848(x^{11/2}+x^{-11/2}) 
+ 
    10604121856(x^{9/2}+x^{-9/2})\\ \nn
&&  + 38955683296(x^{7/2}+x^{-7/2})
 + 
    97377061565(x^{5/2}+x^{-5/2}) \\ \nn
&& + 174547100714(x^{3/2}+x^{-3/2})
+ 
    231970177529(x^{1/2}+x^{-1/2})\,,\eea \bea\nn
H^{(3)}_{15}(x)&=&38(x^{21/2}+x^{-21/2}) + 23524(x^{19/2}+x^{-19/2})\\ \nn
&&  + 2165078(x^{17/2}+x^{-17/2})
 + 
    70882310(x^{15/2}+x^{-15/2})\\ \nn
&&  + 1104236060(x^{13/2}+x^{-13/2})
 + 
    9526203334(x^{11/2}+x^{-11/2})\\ \nn
&&  + 50192855128(x^{9/2}+x^{-9/2}) 
+ 
    173767515352(x^{7/2}+x^{-7/2})\\ \nn
&&  + 419069234222(x^{5/2}+x^{-5/2})
 + 
    736192496108(x^{3/2}+x^{-3/2}) \\ \nn
&& + 969443089694(x^{1/2}+x^{-1/2})\,.
\eea

\subsection*{local $\IF_{2}$} 
By invoking the affine $E_{8}$ Weyl symmetry of local $\frac{1}{2}$ K3 as above, the invariants of $\IF_2$ can be related to those of $\IF_0$ via
\begin{eqnarray}
N^g_{nB+kF}(\IF_0)=N^g_{nB+(k+n)F}(\IF_2) \,.
\end{eqnarray}
Our computations of $\IF_2$ invariants using the large $N$ duality  are in exact agreement with this formula.

\end{document}